\documentclass[11pt]{article}
\pdfoutput = 1

\usepackage[utf8]{inputenc}
\usepackage{color,graphicx}
\usepackage{epsfig}
\usepackage{amsmath}
\usepackage{verbatim}
\usepackage{amssymb}
\usepackage{mathabx}
\usepackage[mathcal]{eucal}
\usepackage{stmaryrd}
\usepackage{empheq}
\usepackage{esint}
\usepackage{physics}
\usepackage{amsfonts}
\usepackage{cite}
\usepackage{array}
\usepackage{setspace}
\usepackage{braket}
\usepackage{float}
\usepackage{url}
\usepackage{mathtools}
\usepackage{tensor}
\usepackage{bbold}
\usepackage{slashed}
\usepackage{tikz}
\usepackage{pgfplots}
\usepackage{graphicx}
\usepackage{epstopdf}
\usepackage{subcaption}
\usepackage[labelfont=bf,font={sf}]{caption}
\usepackage{pgfplots}
\usepackage{mathrsfs}
\usepackage{fancybox}
\usepackage{eurosym}
\usepackage{tcolorbox}
\usepackage{tensor}
\usepackage[normalem]{ulem}
\allowdisplaybreaks
\usepackage{changepage}

\usetikzlibrary{arrows.meta, decorations.markings}

\usepackage[margin = 2.2cm]{geometry}
\usepackage[ragged]{footmisc}
\usepackage[bookmarks=true,bookmarksnumbered=false,
hyperindex=true,bookmarksopen=true,hyperfigures=true,
colorlinks=true,linkcolor=webblue,citecolor=webgreen,urlcolor=webblue,breaklinks]{hyperref}
\definecolor{webgreen}{rgb}{0, 0.5, 0}
\definecolor{webblue}{rgb}{0, 0, 0.5}
\definecolor{webred}{rgb}{0.5, 0, 0}
\definecolor{darkgreen}{rgb}{0,0.5,0}

\def\ben{\begin{equation}}
\def\een{\end{equation}}

   \let\d=\delta 
   
     \let\r=v

\def\be{\begin{equation}}
\def\ee{\end{equation}}
\def\ba{\begin{array}}
\def\ea{\end{array}}

\def\dalemb#1#2{{\vbox{\hrule height .#2pt
       \hbox{\vrule width.#2pt height#1pt \kern#1pt
               \vrule width.#2pt}
       \hrule height.#2pt}}}

\newcommand{\bea}{\begin{eqnarray}}
\newcommand{\eea}{\end{eqnarray}}

\renewcommand{\d}{\mathrm{d}}
\renewcommand{\i}{\mathrm{i}}

\numberwithin{equation}{section}

\begin{document}

\thispagestyle{empty}
    ~\vspace{5mm}
\begin{adjustwidth}{-1cm}{-1cm}
\begin{center}
     {\LARGE \bf 
Sphere amplitudes and observing the universe's size}
   \vspace{0.4in}

     {\bf Andreas Blommaert$^{1}$, Adam Levine$^2$}
     \end{center}
    \end{adjustwidth}
\begin{center}
    \vspace{0.4in}
    {$^1$School of Natural Sciences, Institute for Advanced Study, Princeton, NJ 08540, USA\\
    $^2$Center for Theoretical Physics - a Leinweber Institute, Massachusetts Institute of Technology,\\ Cambridge, MA 02139, USA}
    \vspace{0.1in}
    
    {\tt blommaert@ias.edu, arlevine@mit.edu}
\end{center}

\vspace{0.4in}

\begin{abstract}
\noindent Sine dilaton gravity is holographically related to DSSYK. We explain how to interpret sine dilaton as 2d quantum cosmology. This paves the way for using two copies of DSSYK as hologram for Big-Bang cosmologies. We study the most basic cosmological observable: the sphere amplitude. Via canonical quantization we find a finite answer that matches the on-shell action of a dual matrix integral.

The sphere amplitude (or the norm of the no-boundary wavefunction) also gives a prediction for the universe's size. In the context of slow-roll inflation, the no-boundary state is non-normalizable, and predicts a small universe, in contradiction with experiments. We argue that an avatar of these issues exists in dS JT gravity. By considering sine dilaton as a UV completion of dS JT gravity, the state becomes normalizable. We then consider the observer's no-boundary state and show that this prefers neither small nor large universes. The resulting distribution is flat.
\end{abstract}

\pagebreak
\setcounter{page}{1}
\setcounter{tocdepth}{2}
\tableofcontents

\section{Introduction}
Microscopic holographic descriptions of the early universe are difficult to come by. For asymptotically AdS spacetimes, the holographic duality between SYK at low energies and 2d JT gravity \cite{Maldacena:2016hyu} was crucial in improving our understanding of the gravitational theory \cite{Cotler:2016fpe}. A particular double scaling limit of the SYK model at all energies called DSSYK \cite{Cotler:2016fpe,Berkooz:2018qkz,Berkooz:2018jqr} has often been argued to be related with cosmology \cite{HVerlindetalk,Susskind:2022dfz,Rahman:2022jsf}. A recent series of works \cite{Blommaert:2023wad, Blommaert:2023opb, Blommaert:2024whf, Blommaert:2025avl,Blommaert:2024ydx} has shown that DSSYK is holographically dual to a theory of 2d dilaton gravity called ``sine dilaton'' gravity. The Euclidean action of sine dilaton gravity is\footnote{We are considering units where $L_\text{dS}=1$. Here $\hbar=2\abs{\log q}\hbar_\text{bare}$, with $q$ a parameters of DSSYK \cite{Blommaert:2024ydx} and $\hbar_\text{bare}$ the actual Planck constant (which survives in the JT gravity limit, discussed later). This parameter $\hbar$ appears in all the places where the usual Planck constant would appear. For instance it occurs in the canonical commutation relations.} 
\begin{equation}\label{1.1action}
    I=\frac{1}{2\hbar}\int \mathrm{d} x \sqrt{g}\,(\Phi R+2\sin(\Phi))+\frac{1}{\hbar}\int \d u \sqrt{h}\,\Phi K\,.
\end{equation}
In this work we work out the interpretation of sine dilaton gravity as a theory of 2d quantum cosmology which describes Big-Bang spacetimes.\footnote{Previous work focused on an interpretation of sine dilaton as theory of black holes. Sine dilaton may also be interpreted as holographically related with 3d dS gravity \cite{Narovlansky:2023lfz,Collier:2025lux,Verlinde:2024znh,Tietto:2025oxn}. We will focus on the 2d quantum cosmology interpretation.} This makes the claim that DSSYK is a holographic description of Big-Bang spacetimes precise. The bulk of this work will not concern the relation with DSSYK. We focus instead on sine dilaton quantum cosmology in itself. In the discussion \textbf{section \ref{sect:concl}} we briefly return to the relation with DSSYK and discuss the next steps towards having a working microscopic hologram of 2d Big-Bang cosmologies.

In \textbf{section \ref{sect:2sdcosmo}} we discuss the classical cosmological solutions of sine dilaton gravity. These spacetimes have a Big-Bang and a Big-Crunch. The region near the Big-Bang is described by dS JT gravity \cite{AlHaJe24, Held:2024rmg,Maldacena:2019cbz,Cotler:2019dcj,Cotler:2019nbi,Cotler:2024xzz,Castro:2019vog}. Sine dilaton gravity is a UV completion of dS JT that avoids several UV divergences (as we will indeed see later when we discuss the sphere amplitude).\footnote{The embedding of dS JT gravity was recently nicely discussed in \cite{Okuyama:2025hsd},
 following a shorter discussion in \cite{Blommaert:2024whf}.} We will canonically quantize the theory and compute the exact no-boundary wavefunction \cite{hartle1983wave}
\begin{equation}
    \psi_\text{NB}(\ell,\Phi)=D\int_{-2}^{+2} \d E\,\rho(E)\,\psi_E(\ell,\Phi)=\begin{tikzpicture}[baseline={([yshift=-.5ex]current bounding box.center)}, scale=0.7]
 \pgftext{\includegraphics[scale=1]{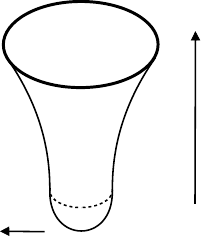}} at (0,0);
    \draw (-3.5,-1.9) node {no boundary};
    \draw (2.45,0.8) node {time};
    \draw (-3.1,1.25) node {fixed $\ell,\Phi$};
  \end{tikzpicture}\label{2.15nobdy}
\end{equation}
Here $\ell$ is the size of the universe, at fixed value of the dilaton $\Phi$ (the ``time'' coordinate):
\begin{equation}
    \ell=\oint\d s\,.
\end{equation}
The expressions for $\psi_E(\ell,\Phi)$ and $\rho(E)$ may be found in \eqref{2.10}, respectively \eqref{2.16rho}. Here, $\rho(E)$ is equal to the spectral density of DSSYK.

In \textbf{section \ref{sect:3spheramplitude}} we compute the sphere amplitude by computing the norm squared of the no-boundary state.\footnote{Recent interesting work on sphere amplitudes includes \cite{Maldacena:2024spf,Ivo:2025yek,Shi:2025amq,Svesko:2022txo,Collier:2025lux,Anninos:2021eit, AnnBar2025}.} The result matches with the prediction from the (plausibly) dual matrix integral \cite{Jafferis:2022wez,Blommaert:2025avl,Saad:2019lba}: 
\begin{equation}
    Z_\text{sphere}=\braket{\psi_\text{NB}\rvert \psi_\text{NB}}=-\int_{-2}^{+2}\d E_1\,\rho(E_1)\int_{-2}^{+2}\d E_2\,\rho(E_2)\,\log\abs{E_1-E_2}\,.\label{1.4sphere}
\end{equation}
This match is universal for rather general periodic dilaton potentials \cite{Blommaert:2024whf} and formally extends also to non-periodic dilaton gravity models like JT gravity \cite{Saad:2019lba} and the $(2,p)$ minimal string \cite{Fan:2021bwt,Mahajan:2021nsd,Mertens:2020hbs,Blommaert:2023wad}. In that case the duality is to a double-scaled matrix integral \cite{Saad:2019lba,Maxfield:2020ale,Witten:2020wvy}. So, the sphere amplitude is divergent, and the wavefunction is non-normalizable. In sine dilaton gravity, the sphere amplitude is instead finite. We will reproduce \eqref{1.4sphere} approximately from a semiclassical calculation, which we interpret as teaching us which solutions one should count when computing sphere amplitudes in general.

In \textbf{section \ref{sect4immortal}} we study the probability distribution of the size $\ell$ for the universe as predicted by the no-boundary state at fixed $\Phi$. For any fixed $\Phi$ (which acts like a time coordinate), the sphere amplitude decomposes as
\begin{equation}
    Z_\text{sphere}=\int_0^\infty \d \ell\rvert_\Phi\,P_\text{sphere}(\ell\rvert_\Phi)\,.\label{1.5sphere}
\end{equation}
In the limit $\hbar \to 0$ with $\Phi=\pi+\hbar \Phi_\text{JT}$, sine dilaton gravity \eqref{1.1action} reduces to dS JT gravity. As explained in section 7.4 of \cite{Blommaert:2024whf} one can deduce this directly at the level of the action. The probability distribution for the size of the universe in dS JT gravity \cite{Maldacena:2019cbz,Cotler:2024xzz,Mahajan:2021nsd,Ivo:2025yek} has avatars of two important issues with the probability distribution of the universe's size as predicted by the Hartle-Hawking no-boundary state in slow-roll inflation \cite{Maldacena:2024uhs,vilenkin1988quantum,Hartle:2007gi,Janssen:2020pii,Lehners:2023yrj,Halliwell:2018ejl}. Firstly, the wavefunction is non-normalizable due to a divergence as $\ell\rvert_\Phi\to 0$. Secondly, the HH wavefunction heavily favors small universes. This is in tension with observations of the spatial curvature of the universe \cite{Planck:2018jri}. We claim by no means to solve these issues in the context of inflation. However, we argue that the avatars of these issues in dS JT gravity \emph{can} be resolved. Firstly, by considering sine dilaton gravity as a UV completion of dS JT gravity, we show that the divergence for $\ell\rvert_\Phi\to 0$ in \eqref{1.5sphere} is resolved. In fact, this distribution vanishes for $\ell\rvert_\Phi\to 0$. This is related with the fact that the sphere amplitude \eqref{1.4sphere} diverges in dS JT gravity but remains finite in sine dilaton gravity. It is not too surprising that a UV completion effectively cut off very short distances, such as $\ell\rvert_\Phi\to 0$. Secondly, we include an observer \cite{Chandrasekaran:2022cip,Witten:2023xze,page1983evolution}. As argued recently in \cite{jonaherez}, the ``observer's no-boundary state'' gets no contribution from the HH sphere geometry. Instead, $\rho_\text{NB}$ is dominated by real-time spacetimes connecting the bra-and the ket boundary conditions:

\begin{align}
\rho_\text{NB}(\ell_1,\Phi_1,q_1|\ell_2,\Phi_2,q_2)=
    \begin{tikzpicture}[baseline={([yshift=-.5ex]current bounding box.center)}, scale=0.7]
 \pgftext{\includegraphics[scale=1]{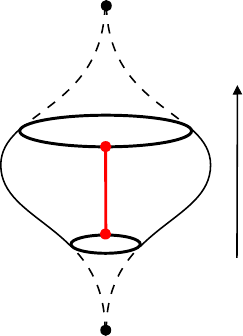}} at (0,0);
    \draw (0.4,-0.35) node {\color{red}obs};
    \draw (2.9,0.8) node {time};
    \draw (-3.7,0.7) node {fixed $\ell_1,\Phi_1$};
    \draw (-0.8,0) node {\color{red}$q_1$};
    \draw (-3,-1.3) node {fixed $\ell_2,\Phi_2$};
    \draw (-0.8,-0.8) node {\color{red}$q_2$};
  \end{tikzpicture}\label{1.6rhointro}
\end{align}
Here $q$ is the observer's energy. This $\rho_\text{NB}$ is the identity matrix on the physical Hilbert space \cite{jonaherez}, which we argue is $L^2(\ell\otimes \Phi)$. So, the observer's no-boundary state does not prefer small nor large universes. It would be interesting to see if such ideas could resolve the issues with the no-boundary state in slow-roll inflation.\footnote{For some recent interesting work in this context see \cite{Fumagalli:2024msi,Betzios:2024oli}.} The fact that describing the state of the universe from an observer's perspective might (one way or another) resolve these issues with the no-boundary state in slow roll was considered in \cite{Hartle:2010dq,Hartle:2010vi}.

In the discussion \textbf{section \ref{sect:concl}}, we briefly return to the relation with DSSYK, and comment critically on assumptions that went into our discussion of the observer's no-boundary state. Some small calculations that clarify statements made in section \ref{sect4immortal} are included as appendices. \textbf{Appendix \ref{app:backgroundsine}} includes background material on sine dilaton gravity.

\section{Sine dilaton quantum cosmology}\label{sect:2sdcosmo}
In this section we present the interpretation of sine dilaton \eqref{1.1action} as a model of 2d quantum cosmology. It can be considered a UV completion of dS JT gravity \cite{AlHaJe24, Held:2024rmg,Maldacena:2019cbz,Cotler:2019dcj,Cotler:2019nbi,Cotler:2024xzz}. In \textbf{section \ref{subsect1.1classicalsolsine}} we discuss the classical solution space, and compare with the solutions in dS JT. In \textbf{section \ref{subsect1.2wavefuncitons}}, we present the exact quantum wavefunctions associated with these solutions. We also pinpoint the linear combination of wavefunctions associated with the no-boundary state \cite{hartle1983wave}.

\subsection{Classical big-bang solutions}\label{subsect1.1classicalsolsine}
The classical solutions of sine dilaton gravity are an example of the general solutions for dilaton gravity models \cite{Gegenberg:1994pv, Witten:2020ert}
\begin{equation}
    \d s^2=F(r)\d \tau^2+\frac{1}{F(r)}\d r^2\,,\quad F(r)=-2\cos(r)+2\cos(\theta)\,,\quad \Phi=r\,.\label{1.9}
\end{equation}
For convenience, we now introduce the parameter $E$ through
\begin{equation}
    E=2\cos(\theta)\,.\label{2.2theta}
\end{equation}
In the black hole interpretation this variable has the interpretation as a holographic ADM energy, and can be interpreted as the eigenvalues of the Hamiltonian in DSSYK \cite{Blommaert:2024ydx}. In this cosmological context, $E$ is not an energy. It is just a parameter labeling classical solutions.\footnote{For comfort of notation in this work we slightly rescaled this energy variable as compared to for instance equation (2.43) in \cite{Blommaert:2025avl} using $E=-2E_\text{DSSYK}$.} The cosmological nature of these solutions \eqref{1.9} becomes apparent when considering complex periods for $\tau$:
\begin{equation}
    \oint \d \tau =\i a\,,
\end{equation}
with real $a$. The classical solutions \eqref{1.9} then read
\begin{equation}
    \d s^2=-\frac{\d \Phi^2}{(2\cos(\Phi)-E)}+a^2(2\cos(\Phi)-E)\,\d x^2\,,\quad x\sim x+1\,,\quad -\text{arccos}(E/2)<\Phi<\text{arccos}(E/2)\,.\label{2.4classicalmetric}
\end{equation}
The parameters $a$ and $E$ label the two-dimensional space of classical solutions of sine dilaton gravity.\footnote{The symplectic structure can be found to be $\sim \d a \wedge \d E$.} It is convenient to introduce the parameter $B$
\begin{equation}
    B=\hbar b=a\sin(\theta)=a\sqrt{1-E^2/4}\,,\label{2.5B}
\end{equation}
which is canonically conjugate to $\theta$ \cite{Blommaert:2025avl}. One more convenient variable is the total volume of a Cauchy slice:
\begin{equation}
    \ell=\oint \d s=a \sqrt{2\cos(\Phi)-E}\,.\label{2.6ell}
\end{equation}
Note that $\Phi$ is a time-like coordinate in \eqref{2.4classicalmetric} and therefore labels different Cauchy slices. The solutions \eqref{2.4classicalmetric} are cosmologies with a Big-Bang and a Big-Crunch. We may picture the universe's size as function of time $\ell(\Phi)$:
\begin{equation}
   \begin{tikzpicture}[baseline={([yshift=-.5ex]current bounding box.center)}, scale=0.7]
 \pgftext{\includegraphics[scale=1]{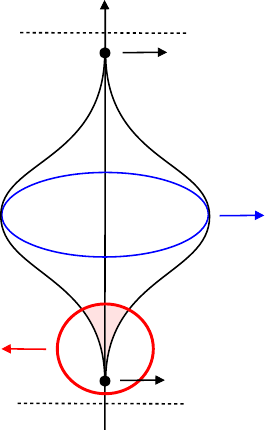}} at (0,0);
\draw (5.8,0) node {\color{blue}maximum size $\ell_\text{max}=\ell\rvert_{\Phi=0}$};
\draw (-5.2,-2.25) node {\color{red}exponential expansion};
\draw (-5.2,-2.95) node {\color{red}dS JT};
\draw (1.9,-2.8) node {big bang};
\draw (2.1,2.75) node {big crunch};
\draw (-2.5,-3.15) node {$-\pi$};
\draw (-2.3,3.1) node {$\pi$};
\draw (2.4,3.5) node {$\Phi$ timelike coordinate};
 
  \end{tikzpicture}\label{2.7plotspacetime}
\end{equation}
The size of the universe vanishes at $\Phi=-\theta$ and at $\Phi=\theta$. Let us describe the evolution of this universe. There is a big-bang singularity at $\Phi=-\theta$, followed by exponential expansion. How long this expansion lasts, depends on the parameter $\theta$. If $\abs{\pi-\theta}\ll 1$ the expansion is approximately exponential for a long time. This is the regime captured by dS JT gravity. Indeed, suppose that $\theta=\pi+\hbar \sqrt{E_\text{JT}}$ and we zoom in on times close to the Big-Bang $\Phi=\pi+\hbar \Phi_\text{JT}$ with $\hbar\to 0$. Then, the metric \eqref{2.4classicalmetric} becomes the usual Big-Bang solution in dS JT gravity:
\begin{align}
    \d s^2&=-\frac{\d \Phi_\text{JT}^2}{(\Phi_\text{JT}^2-E_\text{JT})}+a_\text{JT}^2(\Phi_\text{JT}^2-E_\text{JT})\,\d x^2\,,\quad \Phi_\text{JT}=\sqrt{E_\text{JT}}\cosh(T)>\sqrt{E_\text{JT}}\nonumber\\&=-\d T^2+a_\text{JT}^2 E_\text{JT}\sinh(T)^2\d x^2\label{2.8jt}
\end{align}
Here we rescaled $a=a_\text{JT}/\hbar$. Unlike in dS JT gravity, however, the exponential expansion does not last forever. Indeed, as we see in \eqref{2.7plotspacetime}, the expansion rate slows down until eventually expansion completely stops. At this point, where $\Phi=0$, the universe reaches its maximal size $\ell_\text{max}$. From there, the process is reversed. The universe contracts (slower and slower) until $\Phi=\theta$ where there is a Big-Crunch.

The fact that (exponential) expansion does not last forever and that the universe reaches a maximal size is a universal feature of the dilaton gravity models with a periodic dilaton potential, studied in \cite{Blommaert:2024whf}. These models are UV completions of JT gravity. In the black hole context, they are UV completions in the sense that there is a maximal energy possible for black holes in the system, such that the partition function is finite even at infinite temperature \cite{Blommaert:2024whf}. In the cosmological context, as we will see in section \ref{sect:3spheramplitude}, these are UV completions in the sense that the sphere amplitude is finite, suggesting an underlying finite dimensional Hilbert space. We note that this fact that the partition function is finite for infinite temperature was one of the original motivations for a dS interpretation of DSSYK \cite{HVerlindetalk,Susskind:2022dfz,Rahman:2022jsf}. Therefore we believe this feature should be taken seriously. We briefly comment on how to probe these cosmological spacetimes \eqref{2.4classicalmetric} from the holographic DSSYK description in the discussion section \ref{sect:concl}. A description of cosmological correlators in these spacetimes \eqref{2.4classicalmetric} in terms of SYK variables would realize a microscopic holographic description of 2d Big-Bang cosmologies.

The following is a technical side comment. One can wonder if there are other sectors in the classical solution space aside from these Big-Bang geometries, as in \cite{Held:2024rmg, AlHaJe24}. This is not the case. Physical states in sine dilaton gravity correspond with spacetimes with real values of $E$ and $B$, and therefore also real values of $a$. The fact that physical states have real $B$ was discussed in great detail in section 2 of \cite{Blommaert:2025avl}. It follows from an analysis of the quantum mechanical wavefunctions, which we discuss next.

\subsection{No-boundary wavefunction}\label{subsect1.2wavefuncitons}
We now study the quantum wavefunctions associated with these classical metrics. In 2d dilaton gravity, minisuperspace is exact \cite{henneaux1985quantum,Maldacena:2019cbz,Iliesiu:2020zld,Held:2024rmg}. It consists of the size of the universe $\ell$ \eqref{2.6ell}, the zero mode of $\Phi$, and their conjugates. So it suffices to study wavefunctions $\psi(\ell,\Phi)$. Path integrals prepare wavefunctions that solve the WDW constraint. The WDW constraint can easily be derived in the first order (or ADM) formalism \cite{Held:2024rmg,Iliesiu:2020zld}. In sine dilaton gravity, the constraint is \cite{Blommaert:2025avl}
\begin{equation}
    H_\text{WDW}=\hbar^2\frac{\d}{\d \Phi}\frac{\d}{\d \ell}-\sin(\Phi)\,\ell\,,\quad H_\text{WDW}\,\psi(\ell,\Phi)=0\,.\label{2.9hwdw}
\end{equation}
A complete set of solutions to this differential equation is labeled by the parameter $E$ in \eqref{2.2theta}\footnote{The prefactor is arbitrary. There is no such arbitrary choice of prefactor in the final no-boundary wavefunction \eqref{2.15nobdy}.}
\begin{equation}
    \psi_E(\ell,\Phi)=\frac{\pi^{1/2}}{2^{1/2}}\,H_0^{(1)}(\ell(2\cos(\Phi)-E)^{1/2}/\hbar)\quad \begin{tikzpicture}[baseline={([yshift=-.5ex]current bounding box.center)}, scale=0.7]
 \pgftext{\includegraphics[scale=1]{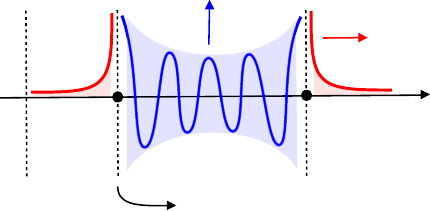}} at (0,0);
\draw (0,2.2) node {\color{blue}oscillations Lorentzian solution};
\draw (3.5,1.15) node {\color{red}decay};
\draw (0.6,-1.7) node {big bang};
\draw (-3.2,-1.5) node {$-\pi$};
\draw (5,-0.3) node {$\Phi$ timelike coordinate};
 
  \end{tikzpicture}\label{2.10}
\end{equation}
The wavefunction behaves as expected of wavefunctions describing the cosmological spacetimes given in \eqref{2.4classicalmetric}. Namely, it oscillates in the region $-\text{arccos}(E)<\Phi<\text{arccos}(E)$ when there is a real time solution, and decays exponentially when the solution is Euclidean. Demanding exponential decay uniquely selects the Hankel function $H_0^{(1)}(x)$ as the normalizable solution of the WDW constraint \eqref{2.9hwdw}.\footnote{In addition \eqref{2.10} has the correct behavior for $\ell\to 0$, where
\begin{equation}
    \psi_E(\ell,\Phi)\supset\frac{1}{\sqrt{2\pi}}\log(2\cos(\Phi)-E)\,.
\end{equation}
This creates a state of approximately fixed dilaton if $\ell\to 0$, which is the Big-Bang (or the horizon, in a black hole context). In other words this state is semiclassically fixing $\Phi_h=\arccos(E/2)$. The following is a technical remark. There is also the $\log(\ell)$ divergent contribution for $\ell\to 0$. This traces back to the contribution from $b=0$ in \eqref{2.11}. This contribution may essentially be ignored, because $b=0$ is a null state as explained in section 2 in \cite{Blommaert:2025avl}. This fixed dilaton state is obtained in the Liouville formulation of sine dilaton gravity \cite{Blommaert:2024ydx,Verlinde:2024zrh,Collier:2025pbm} by inserting the Fourier transform of a bulk vertex operator $V_b$
\begin{equation}
    \ket{E}\quad \leftrightarrow\quad\sum_{b=-\infty}^{+\infty} \cos(b\arccos(E/2))\,V_b\sim \int \d x \sqrt{g}\,e^{-2\pi \Phi/\hbar}\,\delta(\Phi-\arccos(E/2))\,,\quad V_b\sim \int \d x \sqrt{g}\,e^{(\i b-2\pi/\hbar)\Phi }\,.
\end{equation}
}

To derive \eqref{2.10}, one should follow the logic of section 2 of \cite{Iliesiu:2020zld}. Equation \eqref{2.10} differs from equation (2.27) in \cite{Iliesiu:2020zld} by the choice of operator ordering when quantizing $H_\text{WdW}$.\footnote{Their operator ordering would result in exponentials which agree with our wavefunction \eqref{2.10} in the semiclassical limit
\begin{equation}
    \psi_E(\ell,\Phi)=e^{\i \ell (2\cos(\Phi)-E)^{1/2}/\hbar}\,.
\end{equation}} Operator ordering is usually ambiguous in the WdW equation. In sine dilaton gravity, it can be fixed \cite{Blommaert:2025avl} reasonably by demanding equivalence with the quantization of the theory as two copies of Liouville CFT \cite{Blommaert:2024ydx} (such a quantization involves the usual modular data of Liouville CFT \cite{Zamolodchikov:2001ah,Fateev:2000ik,Mertens:2020hbs}). This suggests \eqref{2.9hwdw} is the more natural operator ordering in 2d dilaton gravity.

This wavefunction encodes the classical solution $\ell(E,\Phi)$ \eqref{2.6ell}, as it should. To appreciate this, one can consider the Fourier transform with respect to $\theta=\arccos(E/2)$ (in $E$ variables a Fourier transform involves the Chebychev polynomial $T_b(-E/2)$)\footnote{In the context of a black hole interpretation, wavefunctions of the universe become finite cutoff partition functions  \cite{Iliesiu:2020zld}.}
\begin{equation}
    \psi_b(\ell,\Phi)=\frac{1}{\pi}\int_{-2}^{+2}\frac{\d E}{\sqrt{4-E^2}}T_b(-E/2)\psi_E(\ell,\Phi)\,,\quad B=\hbar b\,.\label{2.11}
\end{equation}
These are precisely the wavefunctions discussed in equation (2.17) of \cite{Blommaert:2025avl}.\footnote{To see this, one uses equation 8.531.1 in \cite{gradshteyn2014table}
\begin{align}
    H_0^{(1)}(\ell \sqrt{2 \cos \Phi - E}/\hbar) = J_0(\ell e^{\i \Phi/2}/\hbar) H^{(1)}_0(\ell e^{-\i \Phi/2}/\hbar) - 2 \sum_{k=1}^\infty J_k(\ell e^{\i \Phi/2}/\hbar) H_k^{(1)}(\ell e^{-\i\Phi/2}/\hbar) \cos (k \theta)\,,E = -2 \cos(\theta)\ .
\end{align}} 
In the semiclassical limit, the argument of the Hankel function becomes large
\begin{equation}
    \psi_E(\ell,\Phi)\to \frac{e^{\i \ell(2\cos(\Phi)-E)^{1/2}/\hbar}}{\ell^{1/2}(2\cos(\Phi)-E)^{1/4}}\,.
\end{equation}
The $E$ integral in \eqref{2.11} is therefore peaked around the classical solution
\begin{equation}
    \frac{B}{\sin(\theta)}=\frac{\ell}{(2\cos(\Phi)-E)^{1/2}}\,.
\end{equation}
Using \eqref{2.5B} we observe that this indeed encodes the semiclassical geometry $\ell(E,\Phi)$ \eqref{2.6ell}, as consistency demanded. Since $\theta$ is periodic, its conjugate $B$ is quantized $B=\hbar b$. A complete set of wavefunctions \eqref{2.11} is spanned by integers $b$ \cite{Blommaert:2025avl}. All physical wavefunctions (with integer $b$) correspond classically with Big-Bang cosmologies. The no-boundary state in dS JT gravity, which we recall is the limit $\hbar\to 0$ of sine dilaton, is instead a geometry with $B=2\pi \i$ \cite{Held:2024rmg}. We now consider the no-boundary wavefunction in sine dilaton gravity, and explain why these two facts are not in contradiction with each other.

Because the no-boundary wavefunction is the result of a gravitational path integral, it satisfies the WDW constraint \eqref{2.9hwdw}. Therefore, it necessarily decomposes into the complete set of solutions $\psi_E(\ell,\Phi)$ to the constraint
\begin{equation}
    \psi_\text{NB}(\ell,\Phi)=D\int_{-2}^{+2} \d E\,\rho(E)\,\psi_E(\ell,\Phi)=\begin{tikzpicture}[baseline={([yshift=-.5ex]current bounding box.center)}, scale=0.7]
 \pgftext{\includegraphics[scale=1]{qcosmo1.pdf}} at (0,0);
    \draw (-3.5,-1.9) node {no boundary};
    \draw (2.45,0.8) node {time};
    \draw (-3.1,1.25) node {fixed $\ell,\Phi$};
  \end{tikzpicture}\label{2.15nobdy}
\end{equation}
The goal is to uniquely determine the value of the expansion coefficients $\rho(E)$. This can be achieved by evaluating the wavefunctions $\psi_E(\ell,\Phi)$ for $\Phi\to \pi/2+\i\infty$, with $\ell$ also diverging so that $\ell e^{-\i \Phi/2}\to +\infty$, and $\ell e^{\i \Phi/2}=\i \beta$ kept finite. These are unusual complex boundary conditions from the point of view of dilaton gravity.\footnote{They are natural however from the point of view of the Liouville formulation of sine dilaton gravity \cite{Blommaert:2024ydx,Verlinde:2024zrh,Collier:2025pbm}. Indeed, they correspond with fixed lengths boundary conditions $L_\text{AdS}=+\infty$, $\bar{L}_\text{AdS}=\i \beta$ \cite{Blommaert:2025avl}. See appendix \ref{app:backgroundsine} for some background material about this. Classically, the state $L_\text{AdS}=+\infty$ is indistinguishable from the FZZT brane with $\mu=\i$ and from the ZZ brane, see \cite{Mertens:2020hbs,Martinec:2003ka}.} However, with these boundary conditions we know that the sine dilaton no-boundary path integral computes the thermal partition function of DSSYK $\psi_\text{NB}(\ell,\Phi)\to Z_\text{DSSYK}(\beta)$ \cite{Blommaert:2024ydx}.\footnote{There is evidence \cite{Blommaert:2025avl} that after including topology change the no-boundary path integral corresponds with computing the expectation value of $\Tr(e^{-\beta H})$ in the q-deformed ETH matrix integral of \cite{Jafferis:2022wez}. Further evidence for this correspondence is provided by matching the sphere amplitudes in sine dilaton and the matrix model in section \ref{sect:3spheramplitude}.} With these boundary conditions (and holographic renormalization), $\psi_E(\ell,\Phi)\to e^{\beta E/2\hbar}$. This is precisely the thermal weight in DSSYK! The expansion coefficient $\rho(E)$ does not depend on the boundary conditions $(\ell,\Phi)$. So, we can uniquely determine $\rho(E)$ by matching $\psi_\text{NB}(\ell,\Phi)$ with the known answer for the sine dilaton disk path integral $Z_\text{DSSYK}(\beta)$ in this limit \cite{Blommaert:2024ydx,Blommaert:2024whf}. This logic follows section 2.4 of \cite{Iliesiu:2020zld}. Hence, we conclude that $\rho(E)$ is exactly the spectral density of DSSYK (or the q-deformed ETH matrix model)
\begin{equation}
   \rho(E)=\sum_{b=-\infty}^\infty \psi(b)\frac{1}{\pi}\frac{T_b(-E/2)}{\sqrt{4-E^2}}\quad\quad  \begin{tikzpicture}[baseline={([yshift=-.5ex]current bounding box.center)}, scale=0.7]
    \pgftext{\includegraphics[scale=1]{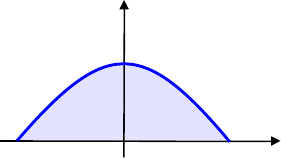}} at (0,0);
    \draw (-2.3,0) node {\color{blue}$\rho(E)$};
    \draw (-2.2,-1.5) node {$-2$};
    \draw (1.5,-1.5) node {$+2$};
    \draw (2.2,-0.5) node {$E$};
  \end{tikzpicture}\label{2.16rho}
\end{equation}

For later purposes we conveniently expanded the DSSYK spectral density in Chebychev polynomials \cite{Okuyama:2023byh}, with the following ``expansion coefficient'':
\begin{equation}
    \psi(b)=\i e^{-\hbar (b+1+2\pi \i /\hbar)^2/8}-\i e^{-\hbar (b+1-2\pi \i /\hbar)^2/8}\,.\label{psib}
\end{equation}
These wavefunctions $\psi(b)$ are the expansion coefficients of the no-boundary state into solutions \eqref{2.11}. The expansion runs over integer $b$. In the dS JT gravity limit, one takes $\hbar\to 0$ whilst keeping $B$ finite. The sum becomes an integral over $B$, and the wavefunction $\psi(b)$ reduces to $\i \delta'(B-2\pi \i)$. Indeed, this is the correct expression in dS JT gravity. It reproduces for instance the Schwarzian path integral \cite{Maldacena:2019cbz} in the asymptotic dS limit. The localization to $B=2\pi \i$ encodes the aforementioned fact that the smooth no-boundary geometry in dS JT gravity is \eqref{2.8jt} with $B\to a\sqrt{E_\text{JT}}=2\pi \i$. We should understand \eqref{psib} as a generalization of this statement to finite $\hbar$. Indeed for small $\hbar$ the wavefunction is a sharp Gaussian ``around'' $B=2\pi \i$.

The salient feature of the function $\rho(E)$ as compared to (for instance) the dS JT gravity limit
\begin{equation}
    \rho(E)\to \rho_\text{JT}(E)\sim \text{sinh}\,(2\pi \sqrt{E_\text{JT}})\,,\label{2.21}
\end{equation}
is that in sine dilaton gravity $\rho(E)$ has compact $E$ support. As may be intuitively obvious, we will find in section \ref{sect:3spheramplitude} that this regulates the UV divergences in the JT gravity sphere amplitude \cite{Maldacena:2019cbz,Mahajan:2021nsd}.\footnote{The sphere in JT gravity is even more divergent than the (2,p) minimal string sphere, with finite p \cite{Mahajan:2021nsd}. This is because the minimal string density of states grows as a power law \cite{Saad:2019lba} whereas the JT one grows exponentially. There is a way of extracting universal data from the sphere amplitude in the minimal string which is independent of how the theory is UV completed. In JT, this universal data is \emph{still} divergent because p is infinite. Our calculations of section \ref{sect:3spheramplitude} can be viewed as fixing the ``non-universal garbage" that Stanford, Mahajan and Yan were not interested in. The sine dilaton calculations can be interpreted as one natural UV completion which fixes that ``garbage''.} Thus, the change of potential $\Phi_\text{JT}\to \sin(\hbar \Phi_\text{JT})$ from dS JT gravity to sine dilaton gravity has the effect of UV completing the dS theory. The fact that including corrections beyond the leading linear behavior in the dilaton gravity potential can result in finite sphere amplitudes was recently independently advocated in \cite{Ivo:2025yek}.

\section{Sphere amplitude}\label{sect:3spheramplitude}
In this section we compute the sphere amplitude in sine dilaton gravity. The methods can be extended to any model of periodic or double-scaled dilaton gravity.\footnote{The extension to general periodic dilaton gravity follows trivially taking into account the discussion of the wavefunctions in periodic dilaton gravity in section 2.3 of \cite{Blommaert:2025avl}.} There is some evidence \cite{Blommaert:2025avl} that sine dilaton gravity is related with a finite-cut matrix integral integral with (leading order) spectrum $\rho(E)$ \eqref{2.16rho}. In \textbf{section \ref{sect3.1matrix}} we compute the matrix model prediction for the sphere amplitude: the on-shell action. In \textbf{section \ref{sect3.1norm}} we reproduce this answer from a gravitational path integral (or canonical quantization) by computing the norm of the no-boundary wavefunction \eqref{2.15nobdy} using an appropriate KG inner product on minisuperspace. The match extends formally to non-periodic dilaton gravity models like JT gravity \cite{Saad:2019lba} and the $(2,p)$ minimal string \cite{Fan:2021bwt,Mahajan:2021nsd,Mertens:2020hbs,Blommaert:2023wad}. In \textbf{section \ref{sect3.3classicalsphere}} we recover (an approximation of) the sphere using semiclassical geometry from a potentially surprising sum over saddles along the lines of \cite{Kruthoff:2024gxc}. This semiclassical decomposition of the exact sphere might be relevant for higher dimensional computations as well.

\subsection{Matrix integral prediction}\label{sect3.1matrix}
There is some evidence that sine dilaton gravity is described by a matrix model \cite{Blommaert:2025avl}. The matrix model partition function is
\begin{equation}
    \mathcal{Z}=\int_{-\infty}^{+\infty}\prod_{i=1}^\text{D}\d \lambda_i\prod_{i<j}(\lambda_i-\lambda_j)^2\exp\bigg(-D\sum_{i=1}^\text{D}V(\lambda_i)\bigg)\,.
\end{equation}
For sine dilaton gravity, $D=2^N$ with $N$ the number of Majorana fermions in SYK. The potential $V(\lambda)$ can be determined from knowledge of the function $\rho(E)$ in \eqref{2.16rho} as follows \cite{DiFrancesco:1993cyw,Eynard:2015aea,Cotler:2016fpe,Mahajan:2021nsd}. Introducing a collective field describing the distribution of the eigenvalues $\lambda_i$
\begin{equation}
    \rho(\lambda)=\frac{1}{D}\sum_{\i=1}^D\delta(\lambda-\lambda_i)\,,
\end{equation}
the matrix integral partition function is rewritten as
\begin{equation}
    \mathcal{Z}=\int \mathcal{D}\rho(\lambda)\dots \exp\bigg( -D^2\int\d \lambda \rho(\lambda) V(\lambda)+D^2\int\d \lambda_1\rho(\lambda_1)\d \lambda_2\rho(\lambda_2)\log \abs{\lambda_1-\lambda_2} \bigg)\,.\label{3.3action}
\end{equation}
Here we left implicit a (complicated) Jacobian. The second integral is a principal value one. The terms in the action are of order $D^2$, hence we can reliably expand around the classical saddle. Variation with respect to $\rho(\lambda)$ determines the potential to be:
\begin{equation}
    V(\lambda)=2\int\d E \rho(E) \log \abs{E-\lambda}\label{3.4potential}
\end{equation}
Taking a $\lambda$ derivative, the resulting equation is a Hilbert transform, which is easily inverted to find the classical solution for $\rho(E)$ given a potential $V(\lambda)$. The potential $V(\lambda)$ which describes the sine dilaton matrix model \cite{Jafferis:2022wez,Okuyama:2023kdo} is a polynomial tuned such that the classical solution of $\rho(E)$ matches with \eqref{2.16rho}.

In the interpretation of matrix integrals as describing 2d quantum gravity \cite{brezin1990exactly,gross1990nonperturbative,douglas1990strings}, $\mathcal{Z}$ represents the path integral over all closed surfaces (not necessarily connected). The free energy of the matrix integral sums therefore over all the connected surfaces
\begin{equation}
    \mathcal{F}=\log \mathcal{Z}=\,\,\begin{tikzpicture}[baseline={([yshift=-.5ex]current bounding box.center)}, scale=0.7]
 \pgftext{\includegraphics[scale=1]{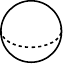}} at (0,0);
  \end{tikzpicture}\,\,+\,\,\begin{tikzpicture}[baseline={([yshift=-.5ex]current bounding box.center)}, scale=0.7]
 \pgftext{\includegraphics[scale=1]{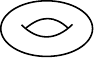}} at (0,0);
  \end{tikzpicture}\,\,+\,\,\dots
\end{equation}
Amplitudes scale as $D^{2-2g}$. The sphere amplitude corresponds with the leading large-$D$ approximation to this free energy, which one obtains from the on-shell action. Inserting \eqref{3.4potential} into the action in \eqref{3.3action}, one obtains the sphere amplitude\footnote{This expression was used for instance to compute the sphere amplitude for the $(2,p)$ minimal string in \cite{Mahajan:2021nsd}.}
\begin{equation}
    \boxed{Z_\text{sphere}=-D^2\int_{-2}^{+2}\d E_1\,\rho(E_1)\int_{-2}^{+2}\d E_2\,\rho(E_2)\,\log\abs{E_1-E_2}\,}\label{3.6sphere}
\end{equation}
In the next section \ref{sect3.1norm} we will show that this matches with the norm of the no-boundary wavefunction \eqref{2.15nobdy}
\begin{equation}
    \boxed{\braket{\psi_\text{NB}\rvert\psi_\text{NB}}=Z_\text{sphere}=\,\, \begin{tikzpicture}[baseline={([yshift=-.5ex]current bounding box.center)}, scale=0.7]
 \pgftext{\includegraphics[scale=1]{qcosmo13.pdf}} at (0,0);
  \end{tikzpicture}}\label{3.7normsquared}
\end{equation}
One can interpreted this match as additional evidence for the identification of sine dilaton gravity with a finite-cut matrix integral \cite{Blommaert:2025avl}.

Notice that \eqref{3.6sphere} will be obviously convergent (recall that $E$ integrals are principal value integrals) for finite-cut matrix integrals and \emph{divergent} for double-scaled matrix integrals. This is consistent with the fact that JT gravity \cite{Maldacena:2019cbz,Cotler:2019nbi} and the $(2,p)$ minimal strings \cite{Mahajan:2021nsd} have UV divergent sphere amplitudes. This is also the case for the Virasoro minimal string \cite{Collier:2023cyw}. Naively one might also think that the matrix model on-shell action would diverge in the complex Liouville string quantization of sine dilaton \cite{Collier:2025pbm}, as this is also a double-scaled (two) matrix integral. However, an intriguing answer was proposed for the sphere amplitude in that model which is finite \cite{Collier:2025lux}.\footnote{This proposal concerned the sphere amplitude in 3d dS. However, using the holographic duality proposed in \cite{Collier:2025lux}, naively this sphere in 3d could be computed holographically by the 2d sphere amplitude in their 2d string theory. Perhaps this is too naive.} It would be interesting to compare that proposal with the on-shell action of the matrix integral and with canonical quantization.

\subsection{Sphere from canonical quantization}\label{sect3.1norm}
We now show how the gravitational path integral together with the correct Klein-Gordon inner product on closed universe reproduces the sphere amplitude \eqref{3.6sphere} prediction of the matrix model. To this end, we compute the norm squared of the no-boundary state. The relevant KG inner product was discussed in \cite{Blommaert:2025avl}. Gauge-fixing to $\ell=0$ slices, this Klein-Gordon inner product reads
\begin{equation}
    \braket{\psi_\text{NB}\rvert \psi_\text{NB}}=\frac{\i}{2}\int_0^{2\pi}\d \Phi\,\bigg(\psi_\text{NB}(\ell=0,\Phi)\frac{\d}{\d \Phi}\psi_\text{NB}(\ell=0,\Phi)^*-\psi_\text{NB}(\ell= 0,\Phi)^*\frac{\d}{\d \Phi}\psi_\text{NB}(\ell=0,\Phi)\bigg)\,.\label{3.8ip}
\end{equation}
Using the decomposition \eqref{2.15nobdy} of the no-boundary wavefunction into the solutions \eqref{2.10} of the WDW constraint:
\begin{equation}
\psi_\text{NB}(\ell,\Phi)=D\int_{-2}^{+2} \d E\,\rho(E)\,\psi_E(\ell,\Phi)\,,    
\end{equation}
we see that the inner product calculation reduces to the following:
\begin{equation}
    \braket{\psi_\text{NB}\rvert \psi_\text{NB}}=D^2\int_{-2}^{+2} \d E_1\, \rho(E_1)\int_{-2}^{+2}\d E_2\,  \rho(E_2)\braket{E_1\rvert E_2}\,.\label{3.10}
\end{equation}
Using the explicit wavefunctions $\psi_E(\ell,\Phi)$ \eqref{2.10} one can compute their inner product as in \eqref{3.8ip}
\begin{equation}
    \braket{E_1\rvert E_2}=-\log\abs{E_1-E_2}\,.\label{3.11ip}
\end{equation}
Inserting this into \eqref{3.10} we find:
\begin{equation}
    \boxed{\braket{\psi_\text{NB}\rvert \psi_\text{NB}}=-D^2\int_{-2}^{+2}\d E_1\,\rho(E_1)\int_{-2}^{+2}\d E_2\,\rho(E_2)\,\log\abs{E_1-E_2}}\label{sphere answer}
\end{equation}
The norm-squared of the no-boundary state, which computes the sphere amplitude in quantum gravity, agrees precisely with the matrix model prediction \eqref{3.6sphere}. Before commenting further on this result, let us take a brief technical detour.

Explicitly performing the $\Phi$ integration to compute the inner product $\braket{E_1\rvert E_2}$ is somewhat subtle. It requires dealing with poles and branchcuts in an appropriate way. Let us present a simpler derivation of the result. Using \eqref{2.16rho} and \eqref{2.11} one finds\footnote{We omitted the contribution from $b=0$ because the associated wavefunction is null \cite{Blommaert:2025avl}.}
\begin{equation}
    \psi_\text{NB}(\ell,\Phi)=D \int_{-2}^{+2}\d E \rho(E)\psi_E(\ell,\Phi)=2 D \sum_{b=1}^\infty \psi(b)\,\psi_b(\ell,\Phi)
\end{equation}
The inner product of the wavefunctions $\psi_b(\ell,\Phi)$ is very simple to calculate, using equation (2.33) in \cite{Blommaert:2025avl} one obtains
\begin{align}
    \braket{b_1|b_2} = \frac{1}{2 b_1}\delta_{b_1 b_2}\,.\label{3.8bortho}
\end{align}
The norm-squared of the no-boundary state is therefore\footnote{Notice that this quantity is positive. It would be interested to understand how this is related with the general expectation of the phase of sphere amplitudes, see recently \cite{Maldacena:2024spf} and references therein.}
\begin{equation}
    \braket{\psi_\text{NB}\rvert\psi_\text{NB}}=2 D^2\sum_{b=1}^\infty \frac{1}{b}\psi(b)^2\,.\label{3.15norm}
\end{equation}
Using the identity
\begin{equation}
    \int_{-2}^{+2}\d E\,\rho(E)T_b(-E/2)=\psi(b)\,,\label{3.16id}
\end{equation}
this can be rewritten as follows
\begin{align}
    \braket{\psi_\text{NB}|\psi_\text{NB}} = 2 D^2\sum_{b=1}^{\infty} \frac{1}{b}\psi(b)^2 =  2 D^2\int_{-2}^{+2} \d E_1\, \rho(E_1)\int_{-2}^{+2}\d E_2\,  \rho(E_2)\  \sum_{b=1}^{\infty} \frac{T_b(-E_1/2) T_b(-E_2/2)}{b}. 
\end{align}
The summation over $b$ can be done using the completeness of the Chebychev polynomials, and results indeed precisely in equation \eqref{sphere answer}.\footnote{This summation is elementary to perform by hand\begin{equation}
    2\sum_{b=1}^\infty\frac{1}{b}\cos(b\,\text{arccos}(E_1/2))\cos(b\text{arccos}(E_2/2))=-\log \abs{E_1-E_2}\,.\label{3.9}
\end{equation}} 

This derivation only relies fundamentally on the identity \eqref{3.16id}, which follows from \eqref{2.16rho}, and the fact that the wavefunctions $\psi_b(\ell,\Phi)$ have inner product \eqref{3.8bortho}. As argued in \cite{Blommaert:2025avl}, $\psi_b(\ell,\Phi)$ is interpreted as a ``trumpet'' amplitude for ending on a circle with boundary conditions determined by $b$, see equation \eqref{bdef}. The no-boundary wavefunction $\psi_\text{NB}$ decomposes into the trumpet and a ``cap'' amplitude equal to $D\,\psi(b)/b$ as follows
\begin{equation}
    \psi_\text{NB}(\ell,\Phi)=\sum_{b=1}^\infty 2b\quad \begin{tikzpicture}[baseline={([yshift=-.5ex]current bounding box.center)}, scale=0.7]
 \pgftext{\includegraphics[scale=.5]{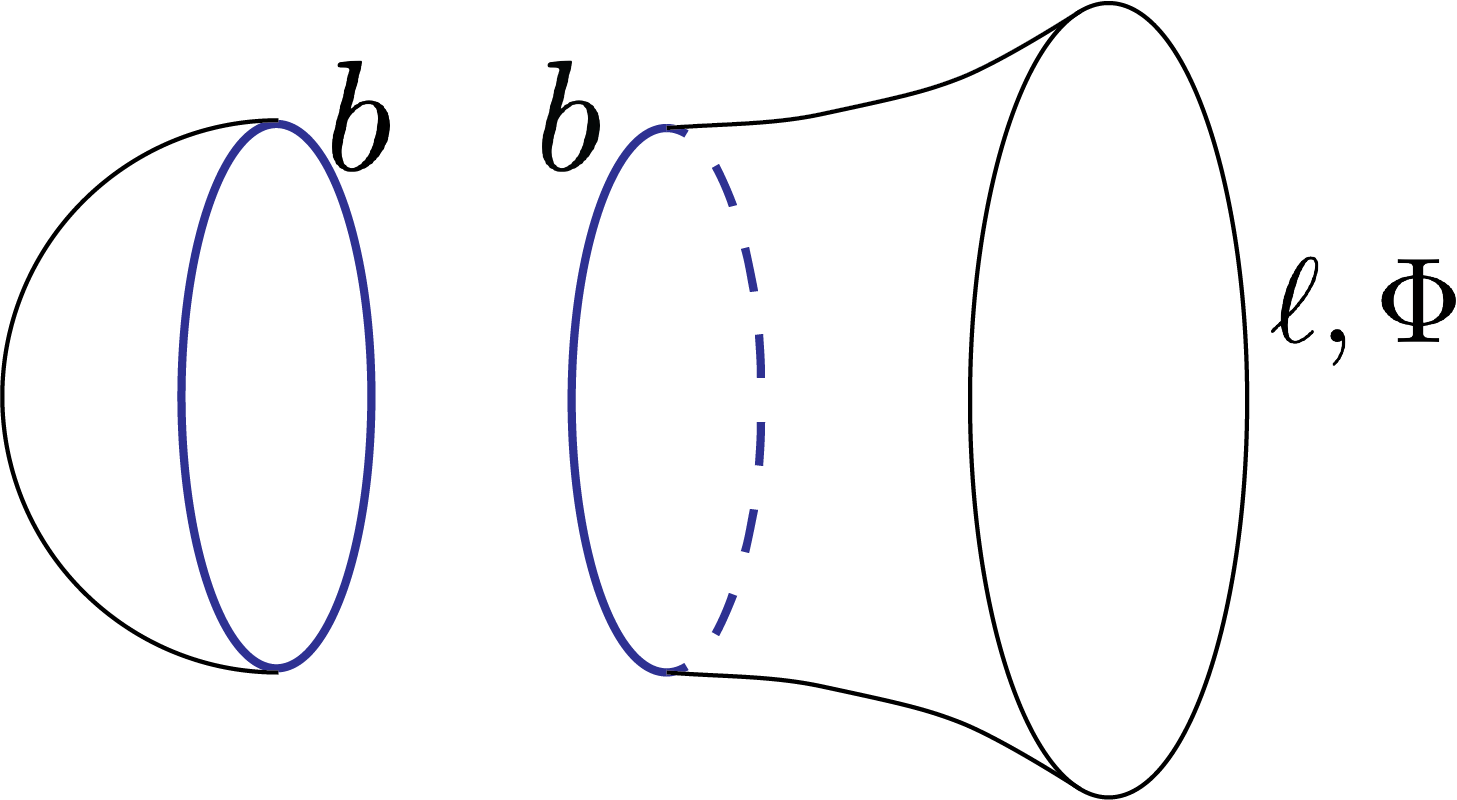}} at (0,0);
  \end{tikzpicture}
\end{equation}
Here, the measure $2b$ follows from inverting the inner product \eqref{3.8bortho} in $b$ basis. This makes the answer \eqref{3.15norm} for the sphere amplitude pictorially obvious. Indeed, one simply glues together two ``caps'' with the correct measure factor $2b$
\begin{equation}
    \braket{\psi_\text{NB}\rvert\psi_\text{NB}}=\sum_{b=1}^\infty 2b\quad \begin{tikzpicture}[baseline={([yshift=-.5ex]current bounding box.center)}, scale=0.7]
 \pgftext{\includegraphics[scale=.5]{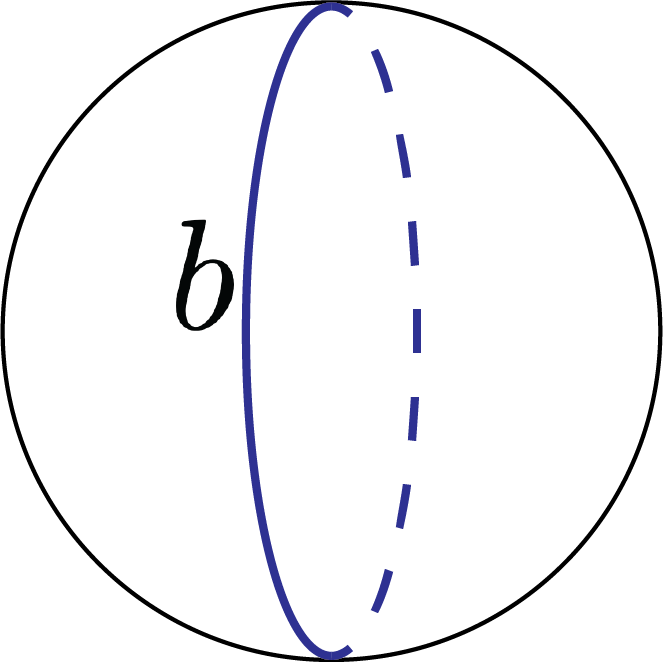}} at (0,0);
  \end{tikzpicture}\label{3.20sphere}
\end{equation}
Both equation \eqref{3.16id} and the fact that the trumpets $\psi_b(\ell,\Phi)$ have the norm \eqref{3.8bortho} was established also for general theories of periodic dilaton gravity in \cite{Blommaert:2025avl}, making our calculation of the sphere amplitude quite general. This provides more evidence for a duality between periodic dilaton gravity theories and finite-cut matrix integrals. 

We stress that according to \eqref{sphere answer}, the no-boundary state for periodic dilaton gravity is \emph{normalizable}. This is to be contrasted with the standard correspondence between double-scaled matrix integrals and dilaton gravity where the UV part of the spectrum grows without bound, such as in JT gravity or the (2,p) minimal string. The best one can hope to do in that context is to match universal contributions to the sphere in gravity and the matrix model, which care only about the low-energy part of the spectrum. This was done in \cite{Mahajan:2021nsd} for the $(2,p)$ minimal string. In our context, both sides are finite and completely match.

\subsection{Semiclassical sphere calculation}\label{sect3.3classicalsphere}
Does the formal answer for the sphere amplitude \eqref{sphere answer} make sense from the point of view of semiclassical gravity? In this section we answer this question affirmatively. We study the classical limit of the exact sphere amplitude, and compare with what the gravitational path integral would semiclasically predict. Our main point can be made in the context of dS JT gravity, which is more familiar terrain than sine dilaton gravity. Therefore, we first consider dS JT and return to sine dilaton at the end of this section.

Recall the classical solutions of dS JT gravity \eqref{2.8jt}
\begin{align}
    \d s^2=-\frac{\d \Phi_\text{JT}^2}{(\Phi_\text{JT}^2-E_\text{JT})}+a_\text{JT}^2(\Phi_\text{JT}^2-E_\text{JT})\,\d x^2\,,\quad E_\text{JT}=\Phi_h^2\,.\label{3.21jtmet}
\end{align}
Furthermore $B=a_\text{JT}\Phi_h$. The classical sphere geometry is found in the interval $-\Phi_h<\Phi_\text{JT}<\Phi_h$. The solution has two horizons (locations where the metric vanishes) at $\Phi=\pm \Phi_h$. The solutions are smooth at both horizons when we tune the free parameter $a_\text{JT}$ such that $B=2\pi \i$. The on-shell action for these geometries is $2\pi$ times the sum of the dilaton values at the two horizons \cite{Maldacena:2019cbz}.\footnote{See also for instance \cite{Svesko:2022txo}.} The free parameter $\Phi_h$ labels the solutions, and the on-shell action is independent of $\Phi_h$.

Let us compare this discussion to the exact sphere amplitude \eqref{sphere answer} with the dS JT density of states \eqref{2.21}. In the classical limit, \emph{ignoring} one loop factors (such as measures), the sphere amplitude reads\footnote{The logarithm was expanded as an integral over $B$ as in \eqref{3.9}. Then we removed the $1/B$ measure factor to reveal the classical structure. This ``equality'' only applies at the level of classical solutions.}
\begin{align}
    Z_\text{sphere}&\overset{\text{classical}}{=} \int_{-\infty}^{+\infty}\d \Phi_{h 1}\,e^{2\pi \Phi_{h1}}\int_{-\infty}^{+\infty}\d \Phi_{h 2}\,e^{2\pi \Phi_{h2}}\,\int_0^\infty \d B\,\cos(B \Phi_{h1})\cos(B \Phi_{h2})\nonumber\\
    &\overset{\text{classical}}{=} \int_{-\infty}^{+\infty}\d \Phi_{h 1}\,e^{2\pi \Phi_{h1}}\int_{-\infty}^{+\infty}\d \Phi_{h 2}\,e^{2\pi \Phi_{h2}}\,(\delta(\Phi_{h1}-\Phi_{h2})+\delta(\Phi_{h1}+\Phi_{h2}))\,.\label{3.22zclas}
\end{align}
The different factors appearing in these equations can be understood as follows. For the first line, one can imagine inserting various Lagrange multiplier fields that fix the dilaton (or area) at the horizon as well as the length $B$ of the geodesic at $\Phi_\text{JT}=0$.\footnote{See for instance \cite{Dong:2018seb,Dong:2022ilf,Blommaert:2023vbz,Cotler:2020lxj,Stanford:2020wkf}.} We then take two half spheres \eqref{3.21jtmet} with generically different values of $\Phi_{h1}$ and $\Phi_{h1}$ and glue them at that geodesic. The on-shell action gets contributions from the conical singularities at the horizons and from the ``kink'' in the geometry at the geodesic. The kink contributes the phase factors $\cos(B \Phi_{h1})\cos(B \Phi_{h2})$, and the singularities contribute $e^{2\pi (\Phi_{h1}+\Phi_{h2})}$. Variation with respect to $B$ selects the geometries without kink at the geodesic. The factor $\delta(\Phi_{h1}+\Phi_{h2})$ on the second line in \eqref{3.22zclas} corresponds with the solution discussed above, where the $\Phi$ contour goes from $-\Phi_h$ to $+\Phi_h$. The resulting action indeed vanishes: $2\pi (\Phi_{h1}+\Phi_{h2})=0$.

The main point that we want to make is that there is a second contribution corresponding with the term $\delta(\Phi_{h1}-\Phi_{h2})$ on the second line of \eqref{3.22zclas}. So, the exact answer for the sphere amplitude suggests that there is another contributing geometry that includes \emph{two} horizons with dilaton value equal to $\Phi_h$. This can be thought of as a complex geometry, defined by a dilaton contour which doubles back on itself in the complex $\Phi$ plane. This contour corresponds with the dashed line below:
\begin{equation}
\begin{tikzpicture}[scale=1]

\coordinate (left) at (-1, 0);
\coordinate (right) at (1, 0);
\coordinate (label) at (3.5,0);

\fill[red] (left) circle (0.08);
\fill[red] (right) circle (0.08);

\node[below left] at (left) {$-\Phi_h$};
\node[below left] at (right) {$+\Phi_h$};

\draw[red, thick, dashed,
      decoration={markings, 
                  mark=at position 0.4 with {\arrow{Stealth}},
                  mark=at position 0.65 with {\arrow{Stealth}}},
      postaction={decorate}]
    (right) .. controls (3, 0.8) and (3, -0.8) .. (right);
\draw[red, thick, decoration = {markings, mark = at position 0.5 with {\arrow{Stealth}}}, postaction = {decorate}] (left) -- (right);
\draw[thick] (2.9,1.1) -- (2.9,.7) -- (3.3,.7);
\node at (3.1,.9) {$\Phi$};
\end{tikzpicture}   
\end{equation}
The solid line denotes the contour for the $\delta(\Phi_{h1}+\Phi_{h2})$ real geometry. In particular, the contour for the complex geometry with both horizons at $\Phi = \Phi_h$ forms a closed loop in dilaton space. One might guess then that the on-shell action for such a contour will be zero. Of course, as we just discussed, there are conical singularities and these will contribute to such a contour at the end points of integration. Thus, the total on-shell action for such a configuration will be given by 
\begin{align}
I_\text{on-shell} = I_\text{bulk} + I_\text{sing} = I_\text{sing}=4\pi \Phi_h\,,
\end{align}
where $I_\text{sing}$ comes from the conical singularities at $\Phi = \Phi_h$. This is indeed the $\delta(\Phi_{h1}-\Phi_{h2})$ contribution in \eqref{3.22zclas}. This discussion extends without significant modifications for a general dilaton gravity dual to a double scaled matrix model. The two horizon points (each) universally contribute $2\pi \Phi_h$, reproducing indeed the semiclassical sphere answer \eqref{sphere answer}.

What about sine dilaton gravity? As discussed in section \ref{sect:2sdcosmo}, the solutions in dilaton gauge take the form \eqref{1.9}
\begin{align}
    \d s^2 = 2(\cos (\Phi) - \cos (\theta))\,a^2 \d x^2 - \frac{\d \Phi^2}{2(\cos(\Phi) - \cos(\theta))}\,\,\quad x\sim x+1\,,\quad \theta=\Phi_h\,.
\end{align}
The first modification as compared to the above dS JT gravity discussion is that there is an additional horizon at $\Phi=2\pi-\theta$.\footnote{By horizon we mean a location where the $x$ component of the metric vanishes. In cosmology, one might interpret this as a beginning  or end of the universe, depending on which coordinate we analytically continue from Euclidean to Lorentzian.} This provides one additional real sphere solution, where the dilaton runs from $\theta\to 2\pi-\theta$. The metric along this contour is in $(-,-)$ signature. Geometries where the sphere-topology metric takes negative definite signature were also considered in the context of the minimal strings \cite{Mahajan:2021nsd}. We can visualize the contour for this additional sphere solution as follows:
\begin{equation}
\begin{tikzpicture}
\begin{axis}[
    axis lines = none,
    xlabel = {$\Phi$},
    ylabel = {},
    domain = -1.57:6.8,
    samples = 300,
    grid = none,
    width = 10cm,
    height = 4cm,
    xmin = -1.57,
    xmax = 6.8,
    ymin = -1.01,
    ymax = 1.01,
    xtick = {-1.047, 1.047, 5.236},
    xticklabels = {$-\theta$, $\theta$, $2\pi-\theta$},
    ytick = {},
    yticklabels = {},
];

\pgfmathsetmacro{\mytheta}{1.047}; 
\pgfmathsetmacro{\costheta}{cos(\mytheta r)}; 

\addplot[blue, thick] {cos(deg(x))};

\addplot[black, thick] coordinates {(-1.57,0) (6.8,0)};
\addplot[red, thick] coordinates {(-\mytheta, \costheta) (\mytheta, \costheta)};
\addplot[red, dashed, thick] coordinates {(\mytheta, \costheta) ({2*pi-\mytheta}, \costheta)};
\addplot[red, mark=*, mark size=2pt] coordinates {(-\mytheta, \costheta)};
\addplot[red, mark=*, mark size=2pt] coordinates {(\mytheta, \costheta)};
\addplot[red, mark=*, mark size=2pt] coordinates {({2*pi-\mytheta}, \costheta)};
\addplot[gray, dashed] coordinates {(-\mytheta, -1.5) (-\mytheta, 1.5)};
\addplot[gray, dashed] coordinates {(\mytheta, -1.5) (\mytheta, 1.5)};
\addplot[gray, dashed] coordinates {({2*pi-\mytheta}, -1.5) ({2*pi-\mytheta}, 1.5)};
\node[blue] at (axis cs: 3.14, -0.7) {$\cos(\Phi)$};
\end{axis}
\end{tikzpicture}
\end{equation}
The solid red-line gives the contour for the positive geometries and the dashed for the $(-,-)$ geometries. One could also imagine contours which wind multiple times around the compact $\Phi$ direction.

The second modification as compared to dS JT gravity is that, in sine dilaton gravity, the on-shell action contribution due to each conical singularity is not given by $e^{2\pi \Phi_h/\hbar}$, but by $e^{2\pi \Phi_h/\hbar-2\Phi_h^2/\hbar}$. This is a well known puzzling feature \cite{Blommaert:2024whf} which we (admittedly) do not understand very well semiclassically. One potential semiclassical interpretation was suggested in \cite{Blommaert:2024ydx}. Namely, counterintuitively it may be that the correct quantum mechanical no-boundary state in sine dilaton looks semiclassically as if there is a ``bare defect, causing the additional Gaussian factor. This seems consistent with the ``fake disk'' interpretation of puzzling features in DSSYK thermodynamics \cite{Lin:2023trc,Lin:2022nss}. For the purposes of this work, this is a side point, which we will not discuss further. For more in depth discussion about this, see \cite{Blommaert:2024whf}.

\section{Observing the universe's size}\label{sect4immortal}
In section \ref{sect:3spheramplitude} we computed the sphere amplitude in 2d dilaton gravity and obtained a reasonable answer. From a cosmological point of view, one is interested in computing observables in the state $\psi_\text{NB}$, rather than knowing the normalization of the state. In this section we consider the simplest ``observable:'' the total size $\ell\rvert_\Phi$ of the universe for some fixed value of the dilaton.

One reason to be interested in this ``observable'' is the superficial relation with an important open problem in cosmology in the context of inflation. As recently reviewed by Maldacena \cite{Maldacena:2024uhs}, Hartle and Hawking's no-boundary wavefunction can be used to predict the size of the universe at a fixed value of the inflaton field $\phi$.\footnote{In the context of inflation, a natural place to evaluate the distribution in $\ell$ is for $\phi$ at its ``reheating value''.} For earlier relevant discussions, see for instance \cite{vilenkin1988quantum,Hartle:2007gi,Janssen:2020pii,Lehners:2023yrj,Halliwell:2018ejl}. The resulting probability distribution is given in equation (12) of \cite{Maldacena:2024uhs}\footnote{In cosmology oftentimes the size of the universe is labeled $a$ instead of $\ell$ where $a$ is the scale factor in the FLRW metric. This is not precisely the same as what we called $a$ in \eqref{2.4classicalmetric}. The correct identification in our notation is with $\ell$.}
\begin{equation}
    P_\text{NB}(\ell\rvert_\phi)\sim \ell\rvert_\phi^{-\text{huge number}}=\begin{tikzpicture}[baseline={([yshift=-.5ex]current bounding box.center)}, scale=0.7]
 \pgftext{\includegraphics[scale=1]{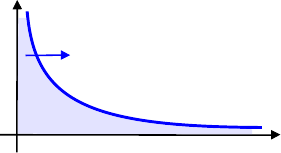}} at (0,0);
    \draw (2.1,0.35) node {\color{blue}small universes preferred};
  \end{tikzpicture}\label{5.1PNBslowroll}
\end{equation}
There are two issues with this distribution
\begin{enumerate}
    \item The distributions heavily pushes the universe to be as small as possible. This is in direct tensions with observations of the spatial curvature of the universe \cite{Planck:2018jri}. The universe is surprisingly large. This problem is usually phrased as a discrepancy in the predicted and measured spatial curvature. However, we find it more intuitive to phrase it as an issue with the predicted \emph{size} of the universe.
    \item It does not make sense to call this a probability distribution, because $P_\text{NB}(\ell\rvert_\phi)$ is not normalizable.
\end{enumerate}

A superficially similar type of behavior is observed in dS JT gravity \cite{Maldacena:2019cbz,Cotler:2024xzz,Mahajan:2021nsd,Ivo:2025yek}. In the holographic regime where
\begin{equation}
    \Phi\to \infty\,,\quad\ell\to \infty\,,\quad L=\frac{\ell}{\Phi}\,\text{ remains finite}\label{4.2ass}
\end{equation}
one obtains a probability distribution $P_\text{NB}(L)$ for the universe's size which like \eqref{5.1PNBslowroll} is not normalizable due to a divergence for $L\to 0$, and which generically favors smaller universes:\footnote{In appendix \ref{app:lengthsJT}, we show how to find this answer from the sphere amplitude along the lines of our calculation in section \ref{sect3.3Pellsphere}.}
\begin{equation}
    P_\text{NB}(L)\sim \frac{1}{L^4}\,.\label{5.3JTdensity}
\end{equation}
The primary goal of this section is to point out that both of these issues can be resolved in sine dilaton gravity. Firstly, in \textbf{section \ref{sect3.3Pellsphere}} by unpacking the sphere amplitude \eqref{sphere answer} we find that $P_\text{NB}(\ell\rvert_\Phi)$ vanishes for $\ell\rvert_\Phi\to 0$ instead of diverging. The result is a normalizable distribution $P_\text{NB}(\ell\rvert_\Phi)$. This had to work, because the sphere amplitude \eqref{sphere answer} is finite. Secondly, in \textbf{section \ref{sect:5.2braketPNB}} we show that a bra-ket wormhole geometry \cite{Chen:2020tes,Fumagalli:2024msi} gives a contribution to $P_\text{NB}(\ell\rvert_\Phi)$ that is exactly \emph{flat} for large values of $\ell\rvert_\Phi$.\footnote{This result is superficially somewhat different from the result of a similar calculation carried out in \cite{Fumagalli:2024msi} who find linear growth instead of a flat distribution. Perhaps this difference can be attributed to the fact that they consider a holographic limit whereas we will consider genuinely finite universes.} However, this contribution is still suppressed by a factor of $1/D^2$ (with $D$ introduced in section \ref{sect3.1matrix}) thus it only competes with the sphere contribution if $\ell\rvert_\Phi\sim D$. We should not be realistically interested in universes of exponential size $\sim D$, so small universes still appear to be favored.

Finally, in \textbf{section \ref{sect:observer}} we include an observer in the closed universe \cite{Chandrasekaran:2022cip,Witten:2023xze,page1983evolution}.\footnote{For related recent work see for instance \cite{Abdalla:2025gzn,Harlow:2025pvj,Akers:2025ahe,Witten:2023xze}.} As explained recently in \cite{jonaherez}, the ``observer's no-boundary state'' gets no contribution from the sphere amplitude. Instead, it is dominated by the bra-ket wormhole. The observer's no-boundary state is exactly the identity matrix on the physical Hilbert space, which we will find to be $L^2(\ell\otimes \Phi)$. So, the observer's no-boundary state favors small nor large universes.\footnote{The bra-ket wormhole results in a non-normalizable distribution. In appendix \ref{app:nonpert} we briefly explain that the distribution becomes normalizable after summing over all geometries, following the same logic as in \cite{Iliesiu:2024cnh,Miyaji:2024ity,Miyaji:2025ucp}.}

We do not claim that this solves the aforementioned issue with the size of the universe in the context of slow roll inflation, but it does suggest that these ideas could perhaps resolve the issue. We list some caveats associated with our present calculations in the discussion, section \ref{sect:concl}.

\subsection{Sphere contribution}\label{sect3.3Pellsphere}
In section \ref{sect3.1norm}, we computed the inner product of solutions to the WdW constraint by gauge-fixing to slices of fixed $\ell\to 0$ in minisuperspace. Alternatively, one may gauge-fix to slices of constant dilaton $\Phi$, resulting in a FP determinant proportional to $\left[ H_\text{WdW},\Phi\right]\sim \d/\d\ell$ \cite{Held:2024rmg,Witten:2022xxp}. This way of rewriting the inner product of the no-boundary wavefunction $\psi_\text{NB}(\ell,\Phi)$ decomposes the sphere amplitude \eqref{sphere answer} as an integral over $\ell$, where the integrand $P_\text{sphere}(\ell\rvert_\Phi)$ is to be interpreted as the (unnormalized) probability that the universe has a size $\ell$ when the dilaton field (the ``clock'') equals $\Phi$
\begin{equation}
    Z_\text{sphere}=\frac{\i}{2}\int_0^{\infty}\d \ell\,\bigg(\psi_\text{NB}(\ell,\Phi)\frac{\d}{\d \ell}\psi_\text{NB}(\ell,\Phi)^*-\psi_\text{NB}(\ell,\Phi)^*\frac{\d}{\d \ell}\psi_\text{NB}(\ell,\Phi)\bigg)=\int_0^\infty \d \ell\,P_\text{sphere}(\ell\rvert_\Phi)\,.\label{3.22zsphere}
\end{equation}
This $P_\text{sphere}(\ell\rvert_\Phi)$ is the appropriate generalization of the usual probability density $\psi(x)\psi(x)^*$ in quantum mechanics.\footnote{We can interpret quantum mechanics as a theory with constraint $H_\text{WDW}=H(p,x)-\i\hbar\, \d/\d T$. Gauge-fixing to constant $T$ leads to a trivial FP determinant, so indeed to integration of $\psi(x)\psi(x)^*$ over some spatial manifold with coordinates $x$ as KG inner product.} Using
\begin{equation}
    \psi_\text{NB}(\ell,\Phi)=\int_{-2}^{+2}\d E \rho(E)\psi_E(\ell,\Phi)=2\sum_{b=1}^\infty \psi(b)\,\psi_b(\ell,\Phi)+\text{null}\,,
\end{equation}
one obtains
\begin{equation}\label{eqn:kgip}
    P_\text{sphere}(\ell\rvert_\Phi) = 2\i\sum_{b_1=1}^\infty\psi(b_1)\sum_{b_2=1}^\infty\psi(b_2)\,\bigg(\psi_{b_1}(\ell,\Phi)\frac{\d}{\d \ell}\psi_{b_2}(\ell,\Phi)^*-\psi_{b_2}(\ell,\Phi)^*\frac{\d}{\d \ell}\psi_{b_1}(\ell,\Phi)\bigg)\,.
\end{equation}
This expression can be plotted numerically. For the sake of a pedagogical presentation, we briefly focus on the diagonal terms with $b_1 = b_2$. We will comment later on the minor effect of the off-diagonal terms with $b_1 \neq b_2$. The resulting expression simplifies further on the slice $\Phi=0$, where the universe reaches it's maximal size: $\ell_\text{max}=\ell\rvert_{\Phi=0}$ \eqref{2.7plotspacetime}. Indeed, inserting the relevant wavefunctions\footnote{Readers may note that the wavefunction \eqref{5.7psi} differs from the one in equation (2.17) of \cite{Blommaert:2025avl} by the addition of a term proportional to $J_b(\ell_{\text{max}}/\hbar)^2$. As discussed in \cite{Blommaert:2025avl}, such a term is orthogonal to all other states using the inner product in \eqref{eqn:kgip}. So we treat it as a ``null state'', the addition of which must be quotient-ed out. Relatedly, the wavefunction in \eqref{5.7psi} for $\Phi \neq 0$ is not manifestly symmetric under $\Phi \to \Phi + 2\pi$ due to a branch cut in $H_b^{(1)}(z)$ on the negative $z$-axis. One can check, however, that the monodromy of $H_b^{(1)}(z)$ is proportional to $J_b(z)$, so the wavefunction in \eqref{5.7psi} shifts by a null state under $\Phi \to \Phi + 2\pi$.}
\begin{equation}
    \psi_b(\ell_\text{max})=\frac{\pi^{1/2}}{2^{1/2}}J_b(\ell_\text{max}/\hbar)H_b^{(1)}(\ell_\text{max}/\hbar)\,,\label{5.7psi}
\end{equation}
and using the \href{https://dlmf.nist.gov/10.5}{Wronskian} of the Hankel functions, one obtains the simpler expression\footnote{Here we rewrote $\psi(b)$ \eqref{psib} in a form which makes more explicit the fact that the wavefunction vanishes on odd integers
\begin{equation}
    \psi(b)=\cos(\pi b/2)e^{-\hbar b^2/8+\hbar b/4}\,,
\end{equation}
The peak of the distribution occurs at $\ell_\text{max}\sim \hbar$. Other features can be studied by interesting readers efficiently by plotting the first few ter,s in this series (convergence is rather rapid).
}
\begin{equation}
    P_\text{sphere}(\ell_\text{max}) \supset \frac{4}{\ell_\text{max}}\sum_{b=1}^\infty e^{-\hbar b(b+1)} J_{2b}(\ell_\text{max}/\hbar)^2\,, \quad\ \begin{tikzpicture}[baseline={([yshift=-.5ex]current bounding box.center)}, scale=0.7]
 \pgftext{\includegraphics[scale=1]{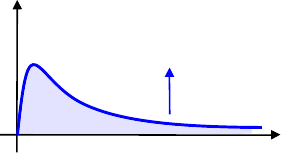}} at (0,0);
     \draw (-1,-1.5) node {$\ell_\text{max} \sim \hbar $};
     \draw (0.6,0.65) node {\color{blue}power-law decay};
  \end{tikzpicture}\label{3.28}
\end{equation}  
Note that, as a consistency check, doing the integral over $\ell_\text{max}$ of the right-hand-side of \eqref{3.28} reproduces the sphere amplitude \eqref{3.20sphere}. The $b_1 \neq b_2$ terms that we ignored in \eqref{3.28} drop out after integrating over $\ell_\text{max}$. Indeed, due to the orthogonality \eqref{3.8bortho}, these off-diagonal terms do not contribute to the sphere amplitude.

We can investigate the density \eqref{3.28} analytically for $\ell_\text{max}\to \infty$ using\footnote{This Heaviside function can be understood also classically using the relation between the parameters $\ell$ and $b$ at $\Phi=0$
\begin{equation}
    \hbar b=\ell \cos(\theta/2)\leq \ell\,.
\end{equation}
One thus expect exponential suppression outside of this region, which indeed happens in the Bessel functions.}
\begin{equation}
    J_{2b}(\ell/\hbar)^2\to \frac{\hbar}{\pi}\frac{1+\sin(2\ell/\hbar)}{\ell}\theta(\ell-2 \hbar b)\,.
\end{equation}
The erratic phase may be ignored for the purposes of discussing the probability distribution. For $\ell_\text{max}\gg \sqrt{\hbar}$, the Gaussian can essentially be summed without bound, resulting in $P_\text{sphere}(\ell_\text{max})\subset 4\sqrt{\hbar}/(\pi \ell_\text{max}^2)$. One can check that after including the off-diagonal contributions with $b_1\neq b_2$ in \eqref{eqn:kgip} that this decay persists: the probability decays quadratically for large universes
\begin{equation}
    P_\text{sphere}(\ell_\Phi)\to  \frac{1}{\ell^2}\,, \text{\,for } \ell\to \infty\,.
\end{equation}
This quadratic decay is universal for all values of $\Phi$. This is therefore also the correct answer in the dS JT gravity regime, where one consider $\Phi=\pi+\hbar \Phi_\text{JT}$.  

The key modification as compared to dS JT gravity happens for $\ell\to 0$, or $\ell_\text{max}\to 0$ in \eqref{3.28}. Sine dilaton gravity can be interpreted as a ``UV completion'' of dS JT gravity. We find that in sine dilaton\footnote{The leading behavior for $\ell\to 0$ comes from $b_1=b_2=1$ in \eqref{eqn:kgip}. For $\Phi=0$ one finds $\sim \ell$, and for generic $\Phi$ one obtains a probability which approaches goes like $\sim -\sin(\Phi)\,\ell \log(\ell)\to 0$.}
\begin{equation}
    P_\text{sphere}(\ell\rvert_\Phi)\to 0\,,\text{\,for } \ell\to 0\,.\label{4.12psphere}
\end{equation}
This resolves the $\ell\to 0$ divergence in dS JT gravity. Even though this is as expected of a UV completion, we still find it reassuring to find this from an exact analytically tractable calculation. More in general, the existence of some non-integrable divergence for $\ell\to 0$ is one-to-one connected (in 2d dilaton gravity) with whether or not the energy integrals in the sphere amplitudes \eqref{sphere answer} converge. For theories with a periodic potential \cite{Blommaert:2024whf}, there is no divergence. But asymptotically AdS theories and minimal strings do have short-distance divergence, and divergent spheres \cite{Maldacena:2019cbz,Cotler:2019nbi,Mahajan:2021nsd,Collier:2023cyw}. The fact that taking into account UV corrections beyond JT gravity could render the sphere amplitude finite was argued independently in \cite{Ivo:2025yek}. We note that, in principle, one would want to study the normalized probability by dividing by the sphere amplitude \eqref{3.22zsphere}, but this changes nothing essential in the previous discussion. The behavior \eqref{4.12psphere} can alternatively be interpreted as a quantum mechanical resolution of the Big-Bang singularity. Indeed, the Big-Bang is characterized by the universe shrinking to a point, thus $\ell\to 0$. In the quantum theory the wavefunction of the universe (or, more precisely, the probability density associated with the wavefunction) has no support on $\ell\to 0$ for any value of the time-like coordinate $\Phi$. Therefore, we never ``encounter'' the Big-Bang.

\subsection{Torus contribution}\label{sect:5.2braketPNB}
In spacetimes which have an asymptotically AdS boundary, calculations in the holographic dual showed us that we are to consider topology change in the gravitational path integral (see for instance \cite{Cotler:2016fpe,Saad:2018bqo}). Without holographic guidance, for closed universes it is unclear if topology change should be considered (and which topologies), even though a natural extension of the AdS/CFT examples suggests to do so. Regardless, there is one contribution to the no-boundary state which should be there in any case. This is the contribution describing spacetimes which are topologically a cylinder connecting the bra-and the ket. Therefore, the no-boundary state of the universe is:
\begin{align}
    \rho_\text{NB}(\ell_1,\Phi_1|\ell_2,\Phi_2)&=D^2\,\psi_\text{NB}(\ell_1,\Phi_2)\psi_\text{NB}(\ell_2,\Phi_2)^*+\rho_\text{cylinder}(\ell_1,\Phi_1|\ell_2,\Phi_2)\,+\text{???} 
\end{align}
This ``cylinder'' contribution is usually imagined as a path integral over complex metrics called bra-ket wormholes \cite{Chen:2020tes}. In the dS JT gravity limit, these complex bra-ket metrics are obtained by taking the big-bang solution \eqref{2.8jt} along the contour $-\infty<T<+\infty$. The big-bang singularity at $T=0$ is avoided by going around it in the complex time plane \cite{Fumagalli:2024msi}, in roughly the same way as the double-cone solution of Saad, Shenker and Stanford (in AdS JT gravity) avoids the singular point at the tip of the cone \cite{Saad:2018bqo}. Schematically, one can picture this ``bra-ket contribution'' as follows

\begin{align}
    \rho_\text{cylinder}(\ell_1,\Phi_1|\ell_2,\Phi_2)&=\begin{tikzpicture}[baseline={([yshift=-.5ex]current bounding box.center)}, scale=0.7]
 \pgftext{\includegraphics[scale=1]{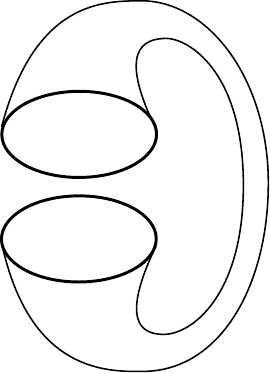}} at (0,0);
    \draw (4.2,0.0) node {complex time};
    \draw (-3.9,0.7) node {fixed $\ell_1,\Phi_1$};
    \draw (-3.9,-1) node {fixed $\ell_2,\Phi_2$};
  \end{tikzpicture}\label{4.15braket}
\end{align}

In this work, we will choose not to interpret this bra-ket wormhole as representing topology change. Indeed, one could pick a time coordinate running from the bra to the ket, such that at every fixed time the spatial topology is a circle. This is more clear when we picture the cylinder as a purely \emph{Lorentzian} spacetime \eqref{2.7plotspacetime} connecting the bra-and the ket:
\begin{align}
    \rho_\text{cylinder}(\ell_1,\Phi_1|\ell_2,\Phi_2)&=\begin{tikzpicture}[baseline={([yshift=-.5ex]current bounding box.center)}, scale=0.7]
 \pgftext{\includegraphics[scale=1]{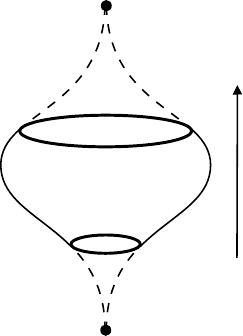}} at (0,0);
    \draw (3.4,0.0) node {real time};
    \draw (-3.6,0.6) node {fixed $\ell_1,\Phi_1$};
    \draw (-2.8,-1.3) node {fixed $\ell_2,\Phi_2$};
  \end{tikzpicture}
\end{align}
Because the spatial topology does not change with time, one can compute this path integral using the first order ADM formulation. As explained in \cite{jonaherez}, for models of 2d dilaton gravity the result is\footnote{The result of this calculation is the same as a path integral over complex metrics \eqref{4.15braket} that were analytically continued from a Euclidean wormhole calculation.}
\begin{equation}
    \rho_\text{cylinder}(\ell_1,\Phi_1|\ell_2,\Phi_2)=\bra{\ell_1.\Phi_1}\frac{1}{2\pi}\int_{-\infty}^{+\infty}\d N\,e^{\i N H_\text{WDW}}\ket{\ell_2,\Phi_2}=\bra{\ell_1.\Phi_1}\mathbb{1}\ket{\ell_2,\Phi_2}\,.\label{5.17rhocyl}
\end{equation}
Here $H_\text{WDW}$ is the WDW Hamiltonian constraint in sine dilaton gravity \eqref{2.9hwdw}. The integration over the Lagrange multiplier $N$ (the timelike component of the metric) results in $\delta(H_\text{WDW})$ which is a projector on the physical Hilbert space spanned by the solutions to the WDW constraint \eqref{2.9hwdw}. Using \eqref{3.8bortho}, we deduce that the identity operator in sine dilaton gravity decomposes as \cite{Blommaert:2025avl}
\begin{equation}
    \mathbb{1}=\sum_{b=1}^\infty 2b\,\ket{b}\bra{b}\,,\quad \braket{b\rvert \ell,\Phi}\equiv\psi_b(\ell,\Phi)\,.
\end{equation}
Therefore
\begin{equation}
    \rho_\text{cylinder}(\ell_1,\Phi_1|\ell_2,\Phi_2)=\sum_{b=1}^\infty 2 b\,\psi_b(\ell_1,\Phi_1)\,\psi_b(\ell_2,\Phi_2)^*\,,
\end{equation}
with the wavefunctions $\psi_b(\ell,\Phi)$ given explicitly by the Bessel functions \eqref{2.11}.

What does does cylinder contribution to the state of the universe predict in terms of the lengths of the universe? Following the discussion of section \ref{sect3.3Pellsphere}, one can find out by decomposing the trace of the cylinder density matrix as follows
\begin{align}
    \Tr(\rho_\text{cylinder})&=\i\sum_{b=1}^\infty b\int_0^{\infty}\d \ell\,\bigg(\psi_b(\ell,\Phi)\frac{\d}{\d \ell}\psi_b(\ell,\Phi)^*-\psi_b(\ell,\Phi)^*\frac{\d}{\d \ell}\psi_b(\ell,\Phi)\bigg)=\int_0^\infty \d \ell\,P_\text{torus}(\ell\rvert_\Phi)\,.
\end{align}
Focusing for analytic simplicity again on the slice $\Phi=0$ and using the same steps that resulted in \eqref{3.28}, one obtains\footnote{Performing the summation yields $\sim \ell_\text{max}(J_0(\ell_\text{max}/\hbar)^2+J_1(\ell_\text{max}/\hbar)^2-J_0(\ell_\text{max}/\hbar)J_1(\ell_\text{max}/\hbar))$. Desired detailed features can be extracted by plotting this quantity using Mathematica. The slope of the linear ``ramp'' is determined by the $b=1$ term, Taylor expanding the Bessel functions. The plateau value is reached for $\ell_\text{max}\sim \hbar$.}
\begin{equation}
    P_\text{torus}(\ell_\text{max})=\frac{1}{\ell_\text{max}}\sum_{b=1}^\infty 2 b\,J_{b}(\ell_\text{max}/\hbar)\qquad \begin{tikzpicture}[baseline={([yshift=-.5ex]current bounding box.center)}, scale=0.7]
 \pgftext{\includegraphics[scale=1]{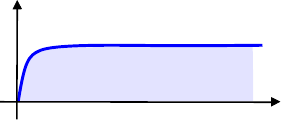}} at (0,0);
    \draw (-1,-1.2) node {$\ell_\text{max}\sim \hbar$};
  \end{tikzpicture}\label{5.21ptorus}
\end{equation}
There are two key features. Firstly, very quickly the density approaches a constant value. The cylinder contribution does not suppress large universes, on the contrary. This resembles the result of \cite{Fumagalli:2024msi}, though they argue that the probability grows linearly with the size of the universe. It would be interesting to understand the source of this discrepancy. The second feature is that this density is not normalizable. This happens because the integrated probability computes the dimension of the physical Hilbert space (ignoring topology change), which is formally infinite
\begin{equation}
    \Tr( \rho_\text{cylinder})=\int_0^\infty \d \ell_\text{max}P_\text{torus}(\ell_\text{max})=\sum_{b=1}^\infty1=\text{dim} (\mathcal{H})\overset{?}{=}Z_\text{torus}=\,\,\begin{tikzpicture}[baseline={([yshift=-.5ex]current bounding box.center)}, scale=0.7]
 \pgftext{\includegraphics[scale=1]{qcosmo17.pdf}} at (0,0);
  \end{tikzpicture}\label{5.22torus}
\end{equation}
It is tempting to interpret this as the torus amplitude in sine dilaton gravity, explaining the title of this section. It would be interesting to understand the torus amplitude in detail. From the matrix integral side one would expect a finite answer. It might be that the above calculation via canonical quantization over-counts tori which in the string theory formulation of sine dilaton gravity correspond with SL$(2,\mathbb{Z})$ images of the torus. Indeed, canonical quantization sometimes ignores mapping class group issues.

Before proceeding, let us remark that it is no surprise that \eqref{5.21ptorus} is approximately flat. Indeed, we know \eqref{5.17rhocyl} that the contribution of the cylinder geometries to the state of the universe is the identity matrix (on the physical Hilbert space). How could the identity operator favor one size of the universe over another? If there were an orthogonal basis of solutions to the WDW constraint associated with a fixed length of the universe, that is an orthogonal basis $\ket{\ell}$, then the distribution \eqref{5.21ptorus} would trivially be \emph{exactly} flat. This is indeed what one finds in the basis $\ket{b}$ where according to \eqref{5.22torus} $P_\text{torus}(b)=1$.\footnote{In practice, the difference shows up in the $\ell$-derivatives that appear in the KG inner product, if there was an orthogonal basis $\ket{\ell}$ at fixed $\Phi$, those derivatives would be absent.} It would be interesting to see if in general dilaton gravity models with periodic potential $P_\text{torus}(\ell_\text{max})\to 1$, perhaps using WKB analysis. We leave this to future work.

\subsection{An observer's perspective}\label{sect:observer}
For large enough sizes of the universe we found in the previous two sections that in sine dilaton gravity
\begin{equation}
    P_\text{sphere}(\ell_\text{max})\sim \frac{D^2}{\ell_\text{max}^2}\,,\quad P_\text{torus}(\ell_\text{max})\sim 1\,.
\end{equation}
This means that the contribution from the cylinder (or bra-ket wormhole) is negligible, unless if $\ell_\text{max}\sim D$. We are not realistically interested in universes whose size is exponential in entropy (at least when it comes to discussing the size of our universe), therefore the bra-ket wormhole contribution seems like it is physically irrelevant. However this drastically changes when considering the no-boundary state from an observer's perspective \cite{jonaherez}.\footnote{Similar ideas outside of the cosmological context were discussed in \cite{Abdalla:2025gzn, Harlow:2025pvj}. Effects of considering an observer were also considered in \cite{Ivo:2024ill}.}

Let us briefly recall the arguments of \cite{jonaherez}. We consider Cauchy slices of sine dilaton gravity with a pointlike observer present, whose mass $q$ is a dynamical parameter
\begin{equation}
    \begin{tikzpicture}[baseline={([yshift=-.5ex]current bounding box.center)}, scale=0.7]
 \pgftext{\includegraphics[scale=1]{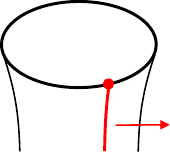}} at (0,0);
    \draw (3.8,-0.8) node {\color{red}pointlike observer};
  \end{tikzpicture}\nonumber
\end{equation}
The usual Hartle-Hawking geometry does not contribute to this observer's no-boundary state. Indeed, the observer's worldline would have no place to smoothly end\footnote{This assumes there is nothing in the theory which couples to the observer in such a way that one could have spontaneous observer creation. This is clearly an approximation. We briefly return to this complaint in the discussion section \ref{sect:concl}.}
\begin{equation}
     \begin{tikzpicture}[baseline={([yshift=-.5ex]current bounding box.center)}, scale=0.7]
 \pgftext{\includegraphics[scale=1]{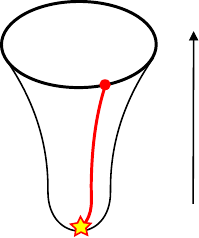}} at (0,0);
    \draw (-0.1,-2.5) node {\color{red}not smooth};
    \draw (2.5,1) node {time};
  \end{tikzpicture}\label{5.24noHH}
\end{equation}
Therefore the leading contribution to the observer's no-boundary state is due to the cylinder (or bra-ket) geometries
\begin{align}
\rho_\text{NB}(\ell_1,\Phi_1,q_1|\ell_2,\Phi_2,q_2)=
    \begin{tikzpicture}[baseline={([yshift=-.5ex]current bounding box.center)}, scale=0.7]
 \pgftext{\includegraphics[scale=1]{qcosmo19.pdf}} at (0,0);
    \draw (0.4,-0.35) node {\color{red}obs};
    \draw (2.9,0.8) node {time};
    \draw (-3.7,0.7) node {fixed $\ell_1,\Phi_1$};
    \draw (-0.8,0) node {\color{red}$q_1$};
    \draw (-3,-1.3) node {fixed $\ell_2,\Phi_2$};
    \draw (-0.8,-0.8) node {\color{red}$q_2$};
  \end{tikzpicture}\label{5.25rhoobspic}
\end{align}
It is possible that one should consider additional contributions from spacetimes with topology change. We discuss the effects from allowing topology change in appendix \ref{app:nonpert}. We remark that in the big-bang cosmologies \eqref{2.4classicalmetric}, a stationary observer has causal access to the full space.\footnote{As explained in \cite{jonaherez} this is determined by calculating the conformal time until the big-bang. The answer diverges, hence an observer has complete causal access.} This rules out the potential option of letting the observer ``circle around'' the cap in \eqref{5.24noHH}, and hiding a second copy of the observer behind a causal horizon (as happens in the preparation for the dS static patch \cite{Chandrasekaran:2022cip}).

We now explicitly compute $\rho_\text{NB}$ for sine dilaton gravity with an observer. Modeling the observer as a point particle with dynamical mass $q$ (following the notation of \cite{Chandrasekaran:2022cip}), the WDW constraint becomes
\begin{equation}
    H_\text{WDW}=H+q\,,\quad H=\hbar^2\frac{\d}{\d \Phi}\frac{\d}{\d \ell}-\ell \sin(\Phi)\,,\quad H_\text{WDW}\,\psi(\ell,\Phi,q)=0\,.\label{4.26wdw}
\end{equation}
Introducing the canonical conjugate variable to the dynamical mass $[q,p]=\i\hbar $, this constraint becomes a Schrodinger equation for a particle in the ``usual'' minisuperspace (with metric $\d s^2=2 \d \Phi\, \d \ell$):
\begin{equation}
    \i \hbar \frac{\d}{\d p}=\hbar^2\frac{\d}{\d \Phi}\frac{\d}{\d \ell}-\ell \sin(\Phi)\,.\label{5.27con}
\end{equation}
Having the constraint be a Schrodinger equation is useful because this simplifies the KG inner product, as explained below equation \eqref{3.22zsphere}. Gauge-fixing to a constant value of $p$ (which plays the role of time) the inner product becomes simply
\begin{equation}
    \braket{\psi_1\rvert\psi_2}=\int\d x\sqrt{g}\,\psi_1(x)\psi_2(x)^*\,.\label{5.28ip}
\end{equation}
Here $x$ are the coordinates on superspace, and the time coordinate $p_0$ of the Cauchy slice to which was gauge-fixed was left implicit in the wavefunctions. A basis of solutions to the constraint \eqref{5.27con} is thus simply the square integrables on minisuperspace of 2d gravity without the observer, which in our case is $L^2(\ell\otimes \Phi)$. Another basis is the coordinate basis (at a fixed time slice), spanned by $\ket{\ell,\Phi}$. The inner product \eqref{5.28ip} has the usual interpretation of inserting a completeness relation in the coordinate basis
\begin{equation}
    \braket{\psi_1\rvert\psi_2}=\int_0^\infty\d \ell\int_{0}^{2\pi}\d \Phi \braket{\psi_1\rvert \ell,\Phi,p_0}\braket{\ell,\Phi,p_0\rvert\psi_2}\,,\quad \braket{\psi_1\rvert \ell,\Phi,p}\equiv\psi_1(\ell,\Phi,p)\,.
\end{equation}
Therefore the identity operator on the physical Hilbert space decomposes as
\begin{equation}
    \mathbb{1}=\int_0^\infty \d \ell\int_0^{2\pi}\d \Phi \ket{\ell,\Phi}\bra{\ell,\Phi}\,.
\end{equation}
As reviewed around \eqref{5.17rhocyl} and explained at length in \cite{jonaherez}, the path integral that computes the observer's no-boundary state \eqref{5.25rhoobspic} computes precisely this identity operator. Therefore
\begin{equation}
    \rho_\text{NB}=\int_0^\infty \d \ell\int_0^{2\pi}\d \Phi \ket{\ell,\Phi}\bra{\ell,\Phi}\,.
\end{equation}
Because now (unlike in section \ref{sect:5.2braketPNB}) there is an orthogonal basis of solutions to the WDW constraint one of the quantum numbers of which is $\ell$, we trivially reach the conclusion that the observer's no-boundary state is a \emph{flat} distribution on the size of the universe
\begin{equation}
    P_\text{NB}(\ell,\Phi)=1\quad\begin{tikzpicture}[baseline={([yshift=-.5ex]current bounding box.center)}, scale=0.7]
 \pgftext{\includegraphics[scale=1]{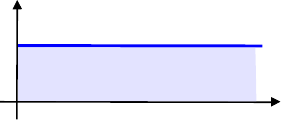}} at (0,0);
     \draw (2,-1.15) node {$\ell$};
  \end{tikzpicture}\label{5.32Pfinal}
\end{equation}
The observer's no-boundary state favors small nor large universes!

Although we only used basic features of quantum mechanics in this subsection, our discussion was perhaps somewhat formal. To remedy this, we include appendix \ref{app:waveJT} where, for JT gravity (for simplicity of presentation), we explicitly find a basis of solutions to the constraint \eqref{5.27con}, and check that there is an orthonormal basis $\ket{\ell,\Phi}$ with unit normalization:
\begin{equation}
    \braket{\ell_1,\Phi_1\rvert \ell_2,\Phi_2}=\delta(\ell_1-\ell_2)\delta(\Phi_1-\Phi_2)\,.\label{4.33main}
\end{equation}

One might wonder, in light of recent discussions of observers in quantum gravity (for instance \cite{Witten:2023xze}), what are the gauge-invariant operators. Gauge-invariant operators commute with the WDW constraint. The operator $\ell$ is not such an operator. The gravitationally dressed (or gauge-invariant) version is
\begin{equation}
    \ell(\mathbf{p}+s)=e^{-\i \mathbf{H}(\mathbf{p}+s)}\,\ell\, e^{\i \mathbf{H}(\mathbf{p}+s)}
\end{equation}
One can then compute expectation values of this in physical states, and obtain the expected answers. For instance\footnote{In the first line, we used the fact that the inner product may be computed using group averaging \cite{Held:2024rmg}. We defined the complete set of states at $p=0$.}
\begin{align}
    \bra{\ell_1,\Phi_1}\ell(\mathbf{p})\ket{\ell_2,\Phi_2}&=\int_{-\infty}^{+\infty}\d N \bra{\ell_1,\Phi_1,p=0}\ell(\mathbf{p})\, e^{\i \mathbf{H} N}\,e^{\i \mathbf{q} N}\ket{\ell_2,\Phi_2,p=0}\nonumber\\
    &=\ell_1\int_{-\infty}^{+\infty}\d N\,\bra{\ell_1,\Phi_2}e^{\i \mathbf{H} N}\ket{\ell_2,\Phi_2}\delta(N)=\ell_1\delta(\ell_1-\ell_2)\delta(\Phi_1-\Phi_2)\,.
\end{align}

The reader might protest that \eqref{5.32Pfinal} is not a normalizable distribution, so that we went backwards from section \ref{sect3.3Pellsphere}. As often, issues with non-normalizability for very large lengths are cured by including topology change, with previous examples including \cite{Iliesiu:2024cnh,Miyaji:2024ity,Miyaji:2025ucp}. As there is no conceptual novelty, nor any (we believe) important physical ramifications, we exile a brief discussion to appendix \ref{app:nonpert}.

\section{Concluding remarks}\label{sect:concl}
In this paper we explained how to interpret sine dilaton gravity as a theory of two-dimensional quantum cosmology. We investigated the no-boundary wavefunction: $\psi_\text{NB}(\ell,\Phi)$. We used the NB wavefunction to compute the sphere amplitude using the inner product $\braket{\psi_\text{NB}\rvert \psi_\text{NB}}$, and recovered the on-shell action of a matrix integral. This works universally for any periodic \cite{Blommaert:2024whf} theory of dilaton gravity. By following an appropriate double-scaling limit, we can also extract UV independent data about the sphere for any non-periodic \cite{Witten:2020wvy,Maxfield:2020ale} theory of 2d dilaton gravity.

As explained in section \ref{subsect1.2wavefuncitons}, in a particular asymptotic double scaling limit of $(\ell,\Phi)$ the no-boundary wavefunction of sine dilaton gravity equals the partition function of DSSYK $\psi_\text{NB}(\beta)=Z_\text{DSSYK}(\beta)$. The equivalence in this asymptotic regime of sine dilaton gravity and DSSYK was shown in \cite{Blommaert:2023wad, Blommaert:2023opb, Blommaert:2024whf, Blommaert:2025avl}. Recent work has also discussed a 3d cosmological interpretation of sine dilaton gravity \cite{Collier:2025lux, Verlinde:2024zrh}. On top of that, several connections between DSSYK and 3d dS gravity have been made \cite{HVerlindetalk,Susskind:2022dfz,Rahman:2022jsf,Narovlansky:2023lfz,Verlinde:2024znh,Tietto:2025oxn,Verlinde:2024zrh,Gaiotto:2024kze}. Here, we discussed the relation between DSSYK and 2d quantum cosmology instead. How all these perspectives are related deserves more detailed investigation.

Have we established a microscopic holographic description of our theory of 2d quantum cosmology? Not quite yet, because the dictionary between cosmological observables in sine dilaton and observables in DSSYK remains to be uncovered. Let us start with the simplest question. How is the sphere encoded in DSSYK? We believe that the finite cutoff no-boundary wavefunction $\psi_\text{NB}(\ell,\Phi)$ corresponds with the partition function of DSSYK with a deformed Hamiltonian, somewhat analogous to $T\bar{T}$ \cite{Harlow:2018tqv,Iliesiu:2020zld}.\footnote{For an investigation in this direction see \cite{Aguilar-Gutierrez:2024oea}.} The sphere amplitude (as computed in section \ref{sect3.1norm}) would thus involve two copies of DSSYK. Two copies also appear in some of the connections between DSSYK and 3d dS physics \cite{Narovlansky:2023lfz,Verlinde:2024zrh}. We intend to investigate these matters in future work.

As an application of our calculation of the sphere and of recent work on the observer's no-boundary state \cite{jonaherez}, we investigated the probability distribution for the size of the universe. We observed that an avatar of the issues with this distribution in the no-boundary state in slow-roll inflation \cite{Maldacena:2024uhs,vilenkin1988quantum,Hartle:2007gi,Janssen:2020pii,Lehners:2023yrj,Halliwell:2018ejl} exists in dS JT gravity. The resulting distribution is non-normalizable because of a divergence for $\ell\to 0$, and favors small universes. We found that the $\ell\to 0$ UV divergence is removed by considering sine dilaton gravity. This could perhaps be viewed as a resolution of the big-bang singularity. We argued that there is no preference for small universes from an observer's perspective. An ``observer's no-boundary state'' is dominated by a bra-ket wormhole topology, and in 2d dilaton gravity favors neither small nor large universes. To end this work we list some critical comments on the interpretation of this result and it's potential relevance for real-world cosmology.
\begin{enumerate}
    \item The resulting distribution is exactly flat. So it can make no meaningful semiclassical predictions (any observable one puts in strongly ``backreacts'' on the distribution). One may hope that in more realistic models there is a semiclassical limit where the distribution is given by the (exponential of the) Euclidean action of the bra-ket wormhole solution \cite{jonaherez}.
    \item The analogy with the no-boundary state in slow-roll is far from perfect. Even in equations. The power law behavior for slow-roll in equation \eqref{5.1PNBslowroll} comes from a Euclidean on-shell action, whereas in 2d dilaton gravity it comes essentially from one-loop determinants and measure factors. In \cite{Ivo:2025yek} it was shown that one could also interpret this as coming from the volume of the moduli space of sphere solutions.
    \item In sine dilaton an observer has complete causal access. In more realistic models of cosmology this is unlikely to be the case. It was argued in \cite{jonaherez} that, even in the case that the observer has causal horizons, the global no-boundary state of the universe is still the identity operator on the global Hilbert space. However, many subtleties arise, not in the least that it is generally unclear how to define gauge invariant observables accessible to the observer in such a scenario.
    \item Treating the observer as an immortal point particle is certainly making mathematical abstraction. Indeed, clearly, real observers are not immortal. It is not clear if there are solid reasons to trust this approximation in this context. However, it seems to give reasonable answers.
\end{enumerate}

\section*{Acknowledgments}
We thank Jordan Cotler, Victor Gorbenko, Oliver Janssen, Jonah Kudler-Flam, Thomas G. Mertens, Jacopo Papalini, Klaas Parmentier, Erez Y. Urbach and Edward Witten for discussions. AB was supported by a Marvin L. Goldberger Membership, the US DOE DE-SC0009988 and by the Ambrose Monell Foundation. AL is supported by the Heising-Simons foundation under grant no. 2023-4430 and the Packard Foundation.

\appendix
\section{Some background material on sine dilaton gravity}\label{app:backgroundsine}
We briefly review the quantization of sine-dilaton gravity discussed in \cite{Blommaert:2023wad, Blommaert:2023opb, Blommaert:2024whf, Blommaert:2025avl}. We begin with the action from the introduction
\begin{align}\label{eqn:sdaction}
I = \frac{1}{2\hbar} \int \d x \sqrt{g} \left( \Phi R + 2 \sin(\Phi) \right) + \frac{1}{\hbar}\int \d u \sqrt{h} \ \Phi K\,,\quad \hbar = 2|\log q|\,.
\end{align}
The boundary terms are appropriate for the standard fixed $\Phi$, fixed length $\ell$ boundary conditions that are familiar from studying JT gravity. The most general boundary conditions are conveniently studied by rewriting sine dilaton classically in terms of two copies of 2d Liouville CFT \cite{Blommaert:2024ydx,Verlinde:2024zrh,Collier:2025pbm}. Writing the metric in conformal gauge $g=e^{2\rho}\hat{g}$, we can define the new fields
\begin{align}\label{eqn:fieldredef}
    \varphi = \rho - \i \Phi/2\,,\quad \bar{\varphi} = \rho + \i \Phi/2\,.
\end{align}
The action \eqref{eqn:sdaction} in terms of $\varphi$ and $\bar{\varphi}$ then indeed becomes the sum of two Liouville actions
\begin{align}
    I_{\text{Liouville}} = \frac{\i}{2\hbar}\int\d x\sqrt{\hat{g}} \bigg((\nabla \varphi )^2 + \hat{R} \varphi + e^{2\varphi }\bigg)\,,\quad I_{\overline{\text{Liouville}}} = -\frac{\i}{2\hbar} \int\d x \sqrt{\hat{g}} \bigg((\nabla \bar{\varphi} )^2 + \hat{R} \bar{\varphi} + e^{2\bar{\varphi} }\bigg)\,.
\end{align}
Here $\hbar=2\abs{\log q}$ is related with the Liouville central charge via $\pi b^2=i\abs{\log q}$. This is the same action as the ``complex Liouville`` string theory discussed in \cite{Collier:2025pbm}. As argued in \cite{Blommaert:2024whf} there are at least two ways to go about quantizing this theory. The quantization of \cite{Collier:2025pbm} is not identical to the quantization that is most closely related with DSSYK. Our main interest is in establishing the sense in which DSSYK is a microscopic hologram of 2d cosmological spacetimes, therefore we stick with the quantization of \cite{Blommaert:2023wad, Blommaert:2023opb, Blommaert:2024whf, Blommaert:2025avl}. It would be interesting to understand the precise relation between the two quantization schemes. This was partially (but far from completely) addressed in \cite{Blommaert:2024whf}, see also section 6 of \cite{Blommaert:2025avl}. Note that to arrive at such a re-writing of the theory, it was important to account for the Gibbons-Hawking-York boundary term in \eqref{eqn:sdaction}.

In Liouville theory, the most natural boundary conditions correspond to fixing the boundary length or the cosmological constant in the \emph{Liouville metric} 
\begin{align}\label{eqn:dilgravliouvillemetric}
   g_L = e^{2\varphi} \hat{g} = e^{-i\Phi} g\,,\quad \overline{g}_L = e^{2\overline{\varphi}} \hat{g} = e^{i\Phi}g\,.
\end{align}
Fixing the cosmological constant corresponds to the standard FZZT brane boundary conditions \cite{Fateev:2000ik}. Using the change of variables in \eqref{eqn:fieldredef}, for instance FZZT$\times$FZZT boundary conditions can be translated to boundary conditions in sine dilaton gravity \cite{Blommaert:2025avl}. Including the boundary terms for the FZZT$\times$FZZT boundary conditions corresponds to a shift of \eqref{eqn:sdaction} by
\begin{align}
    I \to I -\frac{\bar{\mu}}{\hbar}\int \d u \sqrt{h}\  e^{i\Phi/2} -\frac{\mu}{\hbar}\int \d u \sqrt{h} \ e^{-i\Phi/2}.
\end{align}

Classically, the equations of Liouville gravity tell us that the Liouville metrics $g_L$ and $\bar{g}_L$ are locally hyperbolic with curvature $R,\,\bar{R} = -2$. In other words, the classical solutions of sine-dilaton gravity are Weyl equivalent to AdS solutions with the conformal factor $e^{\pm \i \Phi}$. In terms of these AdS metrics, the FZZT brane parameters control the extrinsic curvature of the boundary in each AdS metric, $K_{\text{AdS}}$ and $\overline{K}_{\text{AdS}}$ by the following relation:\footnote{The factors of $i$ are a result of the fact that the conformal transformation between the sine dilaton gravity metric and the AdS/Liouville metric is complex.} 
\begin{align}
    \mu = \i K_{\text{AdS}}\,,\quad \ \overline{\mu} = -\i \overline{K}_{\text{AdS}}\,.
\end{align}
The conjugate to these FZZT parameters are the lengths, $L_{\text{AdS}}$ and $\overline{L}_{\text{AdS}}$, of the boundaries in the two AdS/Liouville metrics. The symplectic form on the phase space of closed universes can then be written as 
\begin{align}\label{eqn:sympformpreconstraint}
    \omega = \d L_{\text{AdS}} \wedge \d K_{\text{AdS}} - \d\overline{L}_{\text{AdS}}\wedge \d\overline{K}_{\text{AdS}}\,.
\end{align}
This is a four dimensional phase space. The physical phase space in 2d dilaton gravity is two-dimensional \cite{Iliesiu:2020zld,henneaux1985quantum,Maldacena:2019cbz,Held:2024rmg}. This happens because the WDW constraint relates the four variables, reducing the phase space to two-dimensions after a symplectic quotient \cite{Blommaert:2025avl}. The WDW constraint in these variables is \cite{Blommaert:2025avl}
\begin{align}
    L_{\text{AdS}}^2 \sqrt{1-K_{\text{AdS}}^2} = B^2 = \overline{L}_{\text{AdS}}^2 \sqrt{1-\overline{K}_{\text{AdS}}^2}\,.\label{bdef}
\end{align}
Here, $B$ is the parameter that we introduced in equation \eqref{2.5B} in the main text. It has the interpretation as the length of the closed geodesic \emph{in the AdS/Liouville metric} homotopic to the FZZT boundary.

One natural set of these boundary conditions then occurs when we take the boundary off to infinity in one of the AdS/Liouville metrics. For example, as in \cite{Blommaert:2025avl}, we can choose $\mu = -\i$ and $\overline{\mu} = \i \cos \theta$. For $\theta$ real, this places the boundary at a location in the ``second'' AdS with extrinsic curvature $|\overline{K}_{\text{AdS}}| <1$. This means that the boundary lies on the double trumpet geometry of AdS \cite{Goel:2020yxl}. One can ask where this geometry lives in the on-shell dilaton gravity metric, $g$. The on-shell solutions of sine-dilaton gravity are \eqref{1.9}. Using \eqref{eqn:dilgravliouvillemetric}, one finds that $\Phi$ is related to the AdS Rindler coordinate $\rho$ by 
\begin{align}
    \Phi = \pi/2 + \i \log \left( \rho + \i \cos(\theta) \right)\,.
\end{align}
The disk geometry then runs from $\rho = \sin \theta$ to $\rho \to + \infty$, which corresponds to taking $\Phi \to \pi/2 + i\infty$. This asymptotic boundary in the AdS/Liouville metric is where the DSSYK hologram lives, as discussed below equation \eqref{2.15nobdy} in the main text.

\section{Distribution of asymptotic lengths in JT}\label{app:lengthsJT}
The purpose of this appendix is to derive equation \eqref{5.3JTdensity} from the main text. The statements is that in dS JT gravity, the distribution of asymptotic lengths $L$ predicted by the sphere amplitude reads
\begin{equation}
    P_\text{NB}(L)\sim \frac{1}{L^4}\,.
\end{equation}
This is equation (82) in \cite{Ivo:2025yek}. Because this is a non-normalizable distribution, the prefactor is irrelevant. Therefore we will only work up to constant prefactors in this appendix. The no-boundary wavefunction in dS JT gravity is
\begin{equation}
    \psi_\text{NB}(\ell,\Phi)=\int_{-\infty}^{+\infty}\d b\,\psi(b)\,\psi_b(\ell,\Phi)\,,
\end{equation}
with trumpet wavefunctions
\begin{equation}
    \psi_b(\ell,\Phi)\sim \frac{1}{\sqrt{-b^2+\ell^2}}e^{\i \Phi\sqrt{-b^2+\ell^2}}\,,\label{bwave}
\end{equation}
and cap amplitude
\begin{equation}
    \psi(b)\sim \delta'(b-2\pi\i)\,.
\end{equation}
We now introduce new coordinates
\begin{equation}
    L=\frac{\ell}{\Phi}\,,\quad \sqrt{\ell \Phi}=X\,.
\end{equation}
The asymptotic limit \eqref{4.2ass} corresponds with taking $X\to\infty$, with fixed and finite $L$. We therefore aim to decompose the KG inner product of the no-boundary state as an integral over $L$, for fixed $X\to \infty$. The metric on superspace transforms as
\begin{equation}
    \d\ell \d \Phi= \d X^2-\frac{1}{4}\frac{X^2}{L^2}\d L^2\,.
\end{equation}
The KG inner product decomposes as
\begin{equation}
    \braket{\psi_\text{NB}\rvert \psi_\text{NB}}=\int_{0}^{\infty}\d L\,P_\text{NB}(L)\,,
\end{equation}
with the probability distribution
\begin{equation}
    P_\text{NB}(L)\sim \int_{-\infty}^{+\infty} \d b_1\, \delta'(b_1-2\pi\i)\int_{-\infty}^{+\infty} \d b_2\, \delta'(b_2-2\pi\i)\frac{X}{L}\psi_{b_1}(X\to\infty,L)\frac{\d}{\d X}\psi_{b_2}(X\to\infty,L)^*
\end{equation}
The wavefunctions \eqref{bwave} become asymptotically
\begin{equation}
    \psi_b(X,L)\sim \frac{1}{X\sqrt{L}}e^{\i X^2}e^{-\i b^2/2L}\,.
\end{equation}
Therefore the probability distribution of asymptotic lengths indeed becomes
\begin{equation}
    P(L)\sim \int_{-\infty}^{+\infty} \d b_1\, \delta'(b_1-2\pi\i)\int_{-\infty}^{+\infty} \d b_2\, \delta'(b_2-2\pi\i)\frac{1}{L^2}e^{\i (b_2^2-b_1^2)\frac{1}{2 L}}\sim \frac{1}{L^4}\,.
\end{equation}

\section{Wavefunction completeness in JT}\label{app:waveJT}
The purpose of this appendix it to demonstrate explicitly that there is an orthonormal basis $\ket{\ell,\Phi}$ for the Hilbert space of closed universes for JT gravity with an observer. Related comments can be found in \cite{jonaherez}. We will consider AdS JT gravity and for simplicity we consider a toy model of the observer \cite{Alexandre:2025rgx} where we replace the pointlike observer by a constant energy density $q=\ell \Lambda$, and quantize that energy density. The WDW Hamiltonian \eqref{4.26wdw} becomes
\begin{equation}
    H_\text{WdW}=\hbar^2\frac{\d}{\d \Phi}\frac{\d}{\d \ell}-\ell(\Phi-\Lambda)\,,\quad [\Lambda,T]=\i \hbar
\end{equation}
This becomes a Schrodinger equation
\begin{equation}
    \i \hbar \frac{\d}{\d T}=\hbar^2\frac{1}{\ell}\frac{\d}{\d\Phi}\frac{\d}{\d \ell}-\Phi\,.\label{cwdw}
\end{equation}
Separation of variables results in the complete set of solutions
\begin{equation}
    \psi_{b,\Lambda}(\ell,\Phi,T)=\frac{1}{\sqrt{2\pi}}\frac{1}{\sqrt{b^2-\ell^2}}e^{\i (\Phi-\Lambda)\sqrt{b^2-\ell^2)}}e^{\i \Lambda T}\,.
\end{equation}
The inner product can be computed using the KG inner product at any slice of fixed $\ell$
\begin{align}
\braket{b_1\rvert b_2}=
    &-\frac{\i}{2} \int_{-\infty}^{+\infty}\d \Phi\int_{-\infty}^{+\infty}\d T \bigg( \psi_{b_1,\Lambda_1}(\ell,\Phi,T)^*\frac{\d}{\d \Phi} \psi_{b_2,\Lambda_2}(\ell,\Phi,T)-\psi_{b_2,\Lambda_2}(\ell,\Phi,T)\frac{\d}{\d \Phi} \psi_{b_1,\Lambda_1}(\ell,\Phi,T)^*\bigg)\nonumber\\
    &=\frac{1}{b_1}\delta(b_1-b_2)\delta(\Lambda_1-\Lambda_2)
\end{align}
To compute this we first performed the $T$ integral. Then we shifted $\Phi-\Lambda_1\to \Phi$ inside the $\Phi$ integration. Alternatively we may compute the inner product at a fixed time slice, as would be the usual procedure in quantum field theory. The wavefunctions at fixed time span a basis of square integrables with measure
\begin{equation}
    \d x\sqrt{g}=\ell\d\ell\d \Phi\,.
\end{equation}
The factor $\ell$ stems from the $1/\ell$ in \eqref{cwdw}. This would be absent for a pointlike observer. It is convenient to introduce the Fourier transformed wavefunctions with respect to $\Lambda$

\begin{equation}
    \psi_{b,P}(\ell,\Phi,T)=\delta(P-T+\sqrt{b^2-\ell^2})\frac{1}{\sqrt{b^2-\ell^2}}e^{\i \Phi\sqrt{b^2-\ell^2}}\,.
\end{equation}
We can explicitly check that these span the square integrables on $\ell\otimes \Phi$
\begin{align}
    &\int_{-\infty}^{+\infty}\d \ell\,\ell\int_{-\infty}^{+\infty}\d \Phi\,\psi_{b_1,P_1}(\ell,\Phi,T_0)\psi_{b_2,P_2}(\ell,\Phi,T_0)^*\nonumber\\
    &= \int_{-\infty}^{+\infty}\d \ell \frac{1}{\sqrt{b_2^2-\ell^2}}\delta(\ell-\sqrt{b_1^2-(P_1-T_0)^2})\,\delta(P_2-T_0+\sqrt{b_2^2-\ell^2})\int_{-\infty}^{+\infty}\d \Phi\, e^{\i \Phi(\sqrt{b_1^2-\ell^2}-\sqrt{b_2^2-\ell^2})}\nonumber\\
    &=\frac{1}{b_1}\delta(b_1-b_2)\int_{-\infty}^{+\infty}\d \ell\,\delta(\ell-\sqrt{b_1^2-(P_1-T_0)^2})\,\delta(P_2-T_0+\sqrt{b_1^2-\ell^2})=\frac{1}{b_1}\delta(b_1-b_2)\delta(P_1-P_2)
\end{align}
Completeness is checked as follows
\begin{align}
    &\int_{-\infty}^{+\infty}\d b\,b\int_{-\infty}^{+\infty}\d P\,\psi_{b,P}(\ell_1,\Phi_1,T_0)\psi_{b,P}(\ell_2,\Phi_2,T_0)^*\nonumber\\
    &=\int_{-\infty}^{+\infty}\d b\,b\int_{-\infty}^{+\infty}\d X\,\delta(X+\sqrt{b^2-\ell_1^2})\delta(X+\sqrt{b^2-\ell_2^2})\frac{1}{\sqrt{b^2-\ell_1^2}\sqrt{b^1-\ell_2^2}}e^{\i \Phi_1\sqrt{b^2-\ell_1^2}}e^{-\i \Phi_2\sqrt{b^2-\ell_2^2}}\nonumber\\
    &=\frac{1}{\ell_1}\delta(\ell_1-\ell_2)\int_{-\infty}^{+\infty}\d b\frac{b}{\sqrt{b^2-\ell_1^2}}e^{\i(\Phi_1-\Phi_2)\sqrt{b^2-\ell_1^2}}=\frac{1}{\ell_1}\delta(\ell_1-\ell_2)\int_{-\infty}^{+\infty}\d a\,e^{\i a (\Phi_1-\Phi_2)}\nonumber\\
    &=\frac{1}{\ell_1}\delta(\ell_1-\ell_2)\delta(\Phi_1-\Phi_2)
\end{align}
In the second line we introduced $X=P-T_0$, and in the third line $a=\sqrt{b^2-\ell_1^2}$. The interpretation of this calculation is the usual fact that the wavefunction at fixed $T=0$ (for instance) may be interpreted as some basis transformation between the basis $\ket{b,P}$ and the coordinate basis $\ket{\ell,\Phi}$
\begin{equation}
    \braket{b,P\rvert\ell,\Phi}=\delta(P-\sqrt{b^2-\ell^2})\frac{1}{\sqrt{b^2-\ell^2}}e^{\i \Phi \sqrt{b^2-\ell^2}}\,.
\end{equation}
The states $\ket{\ell,\Phi}$ defined through these overlaps are orthogonal and complete
\begin{equation}
    \braket{\ell_1,\Phi_1\rvert \ell_2,\Phi_2}=\frac{1}{\ell_1}\delta(\ell_1-\ell_2)\delta(\Phi_1-\Phi_2)\,.
\end{equation}
Asides from the measure factor $1/\ell$ this matches the claimed expression \eqref{4.33main} in the main text. This measure factor traces back to the $1/\ell$ in \eqref{cwdw} and would be absent for pointlike observers.

\section{Including topology change}\label{app:nonpert}
In this appendix we briefly discuss the effects of including topology change on the discussion of section \ref{sect:observer}. Previous very similar discussions can be found in \cite{Iliesiu:2024cnh,Miyaji:2024ity,Miyaji:2025ucp}. See also recently section 3.4 of \cite{jonaherez}. We refer to these works for more detailed explanations. We will consider a basis of asymptotic fixed energy states similar to equation (3.26) in \cite{jonaherez}. After normalization
\begin{equation}
    \left \langle E_1,q_1|E_2,q_2\right\rangle =\delta(E_1-E_2)\delta_{q_1 q_2}\,.
\end{equation}
This equation can be deduced from the cylinder amplitude in sine dilaton gravity, with the matter line inserted along the cylinder. This amplitude is the Laplace transform of equation (3.4) in \cite{Okuyama:2023yat}. One can include topology change by invoking the claimed duality between sine dilaton gravity at the q-deformed ETH matrix model \cite{Blommaert:2025avl}. Assuming such a duality, as in equation (3.43) of \cite{jonaherez}, one would end up (after including corrections from topology change) with some discretized Hilbert space
\begin{equation}
    \left \langle E_i,q_1|E_j,q_2\right\rangle=\delta_{i\,j}\delta_{q_1\,q_2}\,.\label{npid}
\end{equation}
Here the discrete energy values $E_i$ are the eigenvalues of the dual random matrix in one member of the ensemble \cite{Blommaert:2021fob}. The total dimension of this Hilbert space is finite
\begin{equation}
    \dim \mathcal{H}_\text{phys}=2^N\,\dim \mathcal{H}_\text{obs} \nonumber
\end{equation}
In other words the algebra is type $I$ and the observer's no-boundary state is the identity matrix on this finite dimensional Hilbert space. Where does this leave the perturbative basis $\ket{\ell,\Phi}$, and the associated probability distribution of the size of the universe?

This situation is very similar to the discussion of the non-perturbative length of an interval-shaped asymptotically AdS$_2$ universe \cite{Iliesiu:2024cnh}, to which we refer for detailed discussion. Technically speaking, $\ell$ is no longer a physical observable in the non-perturbative theory. This happens because the states $\ket{\ell,\Phi}$ are no longer orthogonal. So, there is no Hermitian operator $\ell$ that acts diagonally on the vectors $\ket{\ell,\Phi}$. To be precise, one could imagine computing the following wavefunctions in the perturbative theory
\begin{equation}
    \left \langle E,q|\ell,\Phi\right\rangle\,.
\end{equation}
The overlaps between the states $\ket{\ell,\Phi}$ could then be computed by inserting an identity operator in the non-perturbative Hilbert space \eqref{npid}
\begin{equation}
    \left \langle \ell_1,\Phi_1|\ell_2,\Phi_2\right\rangle=\sum_{i,a}\left\langle \ell_1,\Phi_1|E_i,q_a\right\rangle\left \langle E_i,q_a|\ell_2,\Phi_2\right\rangle\neq \delta(\ell_1-\ell_2)\delta(\Phi_1-\Phi_2)\,.\label{d4ip}
\end{equation}
Technically this means one can no longer compute the size of the universe. Though it is not clear that this is physically reasonable, one may attempt to define a non-perturbative version of a length operator. It was suggested in \cite{Miyaji:2024ity,Miyaji:2025ucp} to define a discrete and finite dimensional basis of states $\ket{\ell_i,\Phi_a}$ by applying a Gramm-Schmidt procedure on the inner products \eqref{d4ip}. This is more subtle because the original set of vectors $\ket{\ell,\Phi}$ is continuous, but possible. One particularly subtle aspect is that in this GS procedure we have to choose a set of states $\ket{\ell,\Phi}$ which we want to have trustworthy representatives in $\ket{\ell_i,\Phi_a}$ \cite{Miyaji:2025ucp}. Perhaps the most reasonable option is to take a fine mesh of $\Phi$ variables and start GS from $\ell=0$. It is unclear to us how fine this mesh should be taken in $\Phi$ and $\ell$ variables, though a similar question was addressed in \cite{Miyaji:2024ity}. 

The final result is that states $\ket{\ell,\Phi}$ with any value of $\Phi$ and lengths that are sub-exponential in $N$ have accurate representatives in $\ket{\ell_i,\Phi_a}$. If we define a length operator as the matrix that diagonalizes $\ket{\ell_i,\Phi_a}$ with eigenvalues $\ell_i$, then this would approximately measure the size of the universe for universes which are not exponentially large. However, in this construction, one would be definition conclude that there are no universes larger than exponential in $N$. The result of the discussion of section \ref{sect:observer} would be that the observer's no-boundary state predicts a flat distribution on the finite dimensional set $(\ell_i,\Phi_a)$.

We are not quite convinced that the statement that the universe's size cannot exceed exponential in $N$ is correct. The length operator does not approximately measure the universe's size for such large universes. What is true is that universes of size exponential in $N$ can (as states) be expanded in smaller universes. The conclusion at any rate is that non-perturbative effects only play a role in the observer's no-boundary state if one probes exponentially large universes. It seems obvious that we should not be interested in such exponentially large universes. Thus, we conclude that topology change can essentially be ignored, when it comes to predicting the size of the universe using the no-boundary state.

\bibliographystyle{ourbst}
\bibliography{Refs}

\end{document}